\documentclass[preprint,1p,a4paper,11pt]{elsarticle}

\usepackage{amsmath}
\usepackage{booktabs,cellspace}
\usepackage{amssymb,amssymb,mathtools}
\usepackage{url}
\usepackage{xspace}
\usepackage{graphicx}
\usepackage{a4wide}
\usepackage{color}
\usepackage{listings}
\usepackage[absolute,overlay]{textpos}
\usepackage{xcolor}

\usepackage{subcaption}
\usepackage{xstring}
\newcommand{\escapeunderscores}[1]{%
  \StrSubstitute{#1}{_}{\_}%
}

\newcommand{\repolink}[2]{\href{https://github.com/hoppet-code/hoppet/blob/hoppet-2.1.4/#1}{\ttt{#2}}}
\newcommand{\masterlink}[1]{\repolink{#1}{\escapeunderscores{#1}}}

\def\APFELPP{{APFEL\nolinebreak[4]\hspace{-.05em}\raisebox{.4ex}{\tiny\textbf{++}}}}
\DeclareRobustCommand{\CPP}{%
  \mbox{C\nolinebreak[4]\hspace{-.05em}\raisebox{.4ex}{\tiny\textbf{++}}}\xspace
}
\newcommand{\be}{\begin{equation}}
\newcommand{\ee}{\end{equation}}
\newcommand{\bea}{\begin{eqnarray}}
\newcommand{\eea}{\end{eqnarray}}
\newcommand{\bi}{\begin{itemize}}
\newcommand{\ei}{\end{itemize}}
\newcommand{\ben}{\begin{enumerate}}
\newcommand{\een}{\end{enumerate}}

\newcommand{\lp}{\left(}
\newcommand{\rp}{\right)}
\newcommand{\aq}{\alpha_s\left( Q^2 \right)}

\newcommand{\nf}{n_f)}
\newcommand{\nn}{\nonumber}

\newcommand{\GeV}{\;\mathrm{GeV}}
\newcommand{\TeV}{\;\mathrm{TeV}}
\newcommand{\ns}{\;\mathrm{ns}}
\newcommand{\as}{\alpha_s}

\newcommand{\ntlo}{N$^3$LO\xspace}

\definecolor{darkgreen}{rgb}{0,0.6,0}

\newcommand{\eg}{e.g.\ }
\newcommand{\ie}{i.e.\ }

\newcommand{\MSbar}{\overline{\mathrm{MS}}}
\newcommand{\hoppet}{\textsc{hoppet}\xspace}
\newcommand{\ttt}[1]{\texttt{#1}}
\newcommand{\order}[1]{{\cal O}\left(#1\right)}
\newcommand{\fn}{\scriptsize}

\newcommand{\myparagraph}[1]{\paragraph{#1}}

\definecolor{comment}{rgb}{0,0.3,0}
\definecolor{identifier}{rgb}{0.0,0,0.3}

\usepackage{hyperref}
\hypersetup{
  colorlinks = true,
  }

\biboptions{sort&compress}

\newif\ifreleasenote
\releasenotetrue 

\begin{document}
\begin{frontmatter}
\begin{flushright}
  CERN-TH-2023-237\\
  MPP-2023-285\\
  OUTP-23-15P
\end{flushright}
\title{\hoppet{} {\tt v2} release note}

\author[1]{Alexander Karlberg}\ead{alexander.karlberg@cern.ch}
\author[2]{Paolo Nason}\ead{paolo.nason@mib.infn.it}
\author[3,4]{Gavin Salam}\ead{gavin.salam@physics.ox.ac.uk}
\author[5,6]{Giulia Zanderighi}\ead{zanderi@mpp.mpg.de}
\author[7]{Fr\'ed\'eric Dreyer}\ead{dreyer.frederic@gene.com}

\affiliation[1]{organization={CERN, Theoretical Physics Department}, postcode={CH-1211} ,city={Geneva 23}, country={Switzerland}}
\affiliation[2]{organization={INFN, Sezione di Milano-Bicocca, and Universita di Milano-Bicocca, \mbox{Piazza della Scienza 3}}, postcode={20126} ,city={Milano}, country={Italy}}
\affiliation[3]{organization={Rudolf Peierls Centre for Theoretical Physics, Clarendon Laboratory, Parks Road}, postcode={OX1 3PU} ,city={Oxford}, country={UK}}
\affiliation[4]{organization={All Souls College}, postcode={OX1 4AL} ,city={Oxford}, country={UK}}
\affiliation[5]{organization={Max-Planck-Institut fur Physik, Boltzmannstr. 8}, postcode={85748} ,city={Garching}, country={Germany}}
\affiliation[6]{organization={Physik-Department, Technische Universitat Munchen, James-Franck-Strasse 1}, postcode={85748} ,city={Garching}, country={Germany}}
\affiliation[7]{organization={Prescient Design, Genentech, 149 5th Avenue},city={New York}, postcode={NY 10010}, country={USA}}

\begin{abstract}
  We document the three main new features in the v2 release series of the
  \hoppet parton distribution function evolution code, specifically
  support for N$^3$LO QCD evolution in the variable flavour number
  scheme, for the determination of hadronic structure functions for
  massless quarks up to N$^3$LO, and for QED evolution to an accuracy
  phenomenologically equivalent to NNLO QCD.
  Additionally we describe a new Python interface, CMake build option,
  functionality to save a \hoppet table as an LHAPDF grid and update
  our performance benchmarks, including optimisations in
  interpolating PDF tables.
\end{abstract}

\begin{keyword}
  Perturbative QCD \sep DIS \sep DGLAP \sep QED



\end{keyword}
  \begin{textblock*}{2cm}(0.11\textwidth,0.096\textheight)  
    \includegraphics[width=2cm]{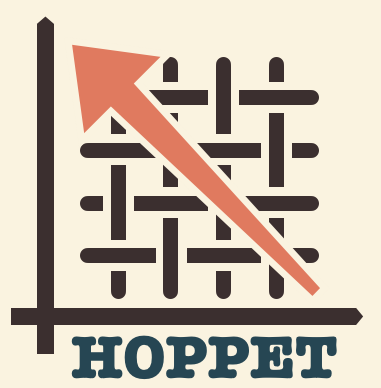}
  \end{textblock*}
\end{frontmatter}
\newpage
\tableofcontents
\section{Introduction}

\hoppet~\cite{Salam:2008qg} is a parton distribution function (PDF)
evolution code written in modern Fortran,  with interfaces also for C/\CPP{} and
earlier dialects of Fortran.
It offers both a high-level PDF evolution interface and user-access to
lower-level functionality for operations such as convolutions of
coefficient functions and PDFs.
It is designed to provide flexible, fast and accurate evolution.

Since the first major release of \hoppet, the landscape of PDF
evolution codes has evolved substantially, with a range of new
open-source codes having been developed, for example the
APFEL~\cite{Bertone:2013vaa}, \APFELPP{}~\cite{Bertone:2017gds} and
EKO~\cite{Candido:2022tld} codes supplementing earlier widely-used
public codes such as QCDNUM~\cite{Botje} and
PEGASUS~\cite{Pegasus}.\footnote{Other codes such as 
  ChiliPDF~\cite{Diehl:2021gvs} appear to not yet be public.}
Nevertheless, \hoppet remains a powerful tool, notably for its ability
to reach high and quantifiable accuracies with competitive speed.
As a result it often provides a critical reference for benchmark
studies, as those presented e.g.\ in
Refs.~\cite{Dittmar:2005ed,Bertone:2024dpm}.
It also provides the option of fast (millisecond-scale) evolution with
accuracies $\sim 10^{-4}$, which is more than sufficient for
phenomenological applications.
Furthermore \hoppet's exposition not just of full PDF
evolution (as exploited in Refs.~\cite{Lai:2010vv,Gao:2013xoa,Butterworth:2015oua,Hou:2019efy,PDF4LHCWorkingGroup:2022cjn}),
but also of low-level functionality though a stable, public API, has
led to its use for a range of
fixed-order~\cite{Caola:2019nzf,Asteriadis:2019dte,Bargiela:2022dla},
all-order~\cite{Banfi:2010xy,Dasgupta:2014yra,Banfi:2015pju,Monni:2016ktx,Bizon:2017rah,Buonocore:2024xmy}
and Monte
Carlo~\cite{Monni:2019whf,vanBeekveld:2023ivn,Buonocore:2024pdv,vanBeekveld:2025lpz}
applications.
With the advent of ever more precise data from the LHC at CERN, and
future very high precision data at the forthcoming Electron-Ion
Collider~\cite{AbdulKhalek:2021gbh} at Brookhaven National Laboratory,
the continued development of tools such as \hoppet remains important
for the field.

This release note documents three major additions to \hoppet, made
available as part of release~2.0.0: 
(1) support for QCD evolution at N$^3$LO in the (zero mass) variable
flavour number scheme (VFNS)~\cite{Buza:1996wv} (Section~\ref{sec:n3lo-evolution});
(2) support for the
determination of massless hadronic structure functions, as initially developed
for calculations of vector-boson fusion, and later deep inelastic scattering, cross
sections~\cite{Cacciari:2015jma,Dreyer:2016oyx,Dreyer:2018qbw,Dreyer:2018rfu,Karlberg:2024hnl}
(Section~\ref{sec:structure-funcs});
(3) support for QED evolution (Section~\ref{sec:qed-evolution}),
originally developed as part of the LuxQED project for the evaluation
of the photon density inside a proton and its extension to lepton
distributions in the
proton~\cite{Manohar:2016nzj,Manohar:2017eqh,Buonocore:2020nai,Buonocore:2021bsf}.

This release also includes a range of other additions relative
to the original 1.1.0 release documented in \cite{Salam:2008qg}, in
particular a Python interface
(Section~\ref{sec:pyinterface}), a CMake-based build system
(Section~\ref{sec:cmake}), and the ability to write LHAPDF grids
(Section~\ref{sec:lhapdf}).
We have also updated the performance benchmarks and improved the speed
for interpolating PDF tables also when loaded from LHAPDF (Section~\ref{sec:v2-performance}).
\hoppet can be obtained by executing
\begin{lstlisting}
  git clone https://github.com/hoppet-code/hoppet.git
  cd hoppet
  git switch -d hoppet-2.x.y  # to switch to a specific release tag, e.g. hoppet-2.1.0
\end{lstlisting}
Unified documentation of the whole \hoppet package is part of the
distribution at \url{https://github.com/hoppet-code/hoppet} in the
\masterlink{docs/manual} directory.
Details of the other changes since release 1.1.0 can be found in the
\masterlink{NEWS.md} and \masterlink{ChangeLog} files from the
repository.
\medskip

\noindent \fbox{\parbox{\textwidth}{
\textbf{Note that as of the 2.0.0 release, \hoppet's library name has been renamed
from \texttt{hoppet\_v1} to \texttt{hoppet}, and similarly for the main
module and \CPP include file.
Users with an existing installation of \hoppet (v1) should make sure
that they link with the new library name.
Apart from this, v2 remains fully backwards compatible with code
designed for v1.
}
}}

\section{Perturbative evolution in QCD}
\label{sec:pqcd}

\ifreleasenote%
  \textbf{Note:} this section is largely a repetition of Section~2 of
  Ref.~\cite{Salam:2008qg}, and is included again here to help provide
  context for the discussion that follows in subsequent
  sections.\medskip
\fi

First of all we set up the notation and
conventions that are used throughout \hoppet. The DGLAP
equation for a non-singlet parton distribution reads
\begin{equation}
  \label{eq:dglap-ns}
  \frac{\partial q(x,Q^2)}{\partial \ln Q^2} = 
\frac{\aq}{2\pi}\int_x^1 \frac{dz}{z}
  P(z,\aq) q\lp \frac{x}{z},Q^2\rp \equiv 
\frac{\aq}{2\pi}  P(x,\aq) \otimes q\lp x,Q^2\rp \ .
\end{equation}
The related variable $t\equiv \ln Q^2$ is also used
in various places in \hoppet.
The splitting functions in eq.~(\ref{eq:dglap-ns})
are known exactly up to NNLO in the 
unpolarised case \cite{Furmanski:1980cm,Curci:1980uw,NNLO-NS,NNLO-singlet}, and approximately at N$^3$LO~\cite{Gracey:1994nn,Davies:2016jie,Moch:2017uml,Gehrmann:2023cqm,Falcioni:2023tzp,Gehrmann:2023iah,McGowan:2022nag,NNPDF:2024nan,Moch:2021qrk,Falcioni:2023luc,Falcioni:2023vqq,Moch:2023tdj,Falcioni:2024xyt,Falcioni:2024qpd}:
\begin{equation}
  \label{eq:dpdf}
   P(z,\aq)=P^{(0)}(z)+\frac{\aq}{2\pi}P^{(1)}(z)+
\lp \frac{\aq}{2\pi} \rp^2 P^{(2)}(z) 
+
\lp \frac{\aq}{2\pi} \rp^3 P^{(3)}(z) \ ,
\end{equation}
and up to NNLO \cite{Mertig:1995ny,Vogelsang:1996im,Moch:2014sna,Moch:2015usa,Blumlein:2021enk,Blumlein:2021ryt} in the polarised case.
The generalisation to the singlet case is straightforward, as it
is 
to the case of time-like evolution\footnote{
The general structure of the relation between space-like
and time-like evolution and splitting functions
 has been investigated in
 \cite{Furmanski:1980cm,Curci:1980uw,Stratmann:1996hn,Dokshitzer:2005bf,Mitov:2006ic,Basso:2006nk,Dokshitzer:2006nm,Beccaria:2007bb}.
 See references to those articles for more recent updates.
}, 
relevant for example for fragmentation function analysis,
where NNLO results
are also available \cite{Mitov:2006ic,Moch:2007tx,Almasy:2011eq}.

As with the splitting functions, most perturbative quantities in
\hoppet are defined to be coefficients of powers of $\as/2\pi$.
However there are some places where a different convention is used,
either for historical reasons or because external code uses a
different convention.
In particular the $\beta$-function coefficients of the running
coupling equation,
\begin{equation}
  \label{eq:as-ev}
  \frac{d\as}{d\ln Q^2} = \beta\lp \aq\rp = -\as (\beta_0\as +
  \beta_1\as^2 + 
  \beta_2\as^3 + 
  \beta_3\as^4) \,,
\end{equation}
are defined internally in \hoppet as multiplying powers of $\alpha_s$
directly.

The evolution of the strong coupling and the parton distributions can
be performed in both the fixed flavour-number scheme (FFNS) and the 
variable flavour-number scheme (VFNS). In the VFNS case we 
need the matching conditions between the effective
theories with $n_f$ and $n_{f}+1$ light flavours for both the strong 
coupling $\aq$ and the parton distributions at the heavy quark
mass threshold $m_h^2$.

These matching conditions for the parton distributions
receive non-trivial contributions at higher orders. In the $\MSbar$
(factorisation) scheme, for example,
carrying out the matching at a scale equal to the heavy-quark mass,
these begin at NNLO:\footnote{In
  a general scheme they would start at NLO.} %
for light quarks $q_{l,i}$ of flavour $i$ 
(quarks that are considered massless
below the heavy quark mass threshold $m_h^2$) the matching between
their values in the $n_f$ and
$n_f+1$ effective theories reads\footnote{Note that the literature
  focuses on the $q_{l,i}+q_{l,-i}$ combination. We thank Johannes
  Bl\"umlein for having provided the $A^{\rm NS,-}_{qq,h}(x)$ code
  needed for the $q_{l,i}-q_{l,-i}$ combination.}:
\begin{align}
\label{eq:lp-nf1}
  q_{l,i}^{\,(n_f+1)}(x,m_h^2) + q_{l,-i}^{\,(n_f+1)}(x,m_h^2)  & =   q_{l,i}^{\,(\nf}(x,m_h^2) + q_{l,-i}^{\,(\nf}(x,m_h^2) \notag \\ &+
   A^{\rm NS,+}_{qq,h}(x) \otimes \left(
   q_{l,i}^{\, (\nf}(x,m_h^2) + q_{l,-i}^{\, (\nf}(x,m_h^2)\right) \notag\\
   & + \frac{1}{n_f} \Big\{A^{\rm PS}_{qq,h}(x) \otimes \Sigma^{\, (\nf}(x,m_h^2) \notag\\
   & + A^{\rm S}_{qg,h}(x) \otimes g^{\, (\nf}(x,m_h^2)\Big\} \ , \notag \\
  q_{l,i}^{\,(n_f+1)}(x,m_h^2) - q_{l,-i}^{\,(n_f+1)}(x,m_h^2)  & =   q_{l,i}^{\,(\nf}(x,m_h^2) - q_{l,-i}^{\,(\nf}(x,m_h^2) \notag \\ &+
   A^{\rm NS,-}_{qq,h}(x) \otimes \left(
   q_{l,i}^{\, (\nf}(x,m_h^2) - q_{l,-i}^{\, (\nf}(x,m_h^2)\right) \ ,
\end{align}
where $i = 1,\ldots n_f$, while for the gluon distribution, the heavy
quark PDF $q_h$, and the singlet PDF $\Sigma(x,Q^2)$ (defined in
Table~\ref{eq:diag_split}) one has :
\begin{align}
\label{eq:hp-nf1}
  g^{(n_f+1)}(x,m_h^2)  &=
    g^{\, (\nf}(x,m_h^2) +
    A_{\rm gq,h}^{\rm S}(x) \otimes \Sigma^{(\nf}(x,m_h^2) +
    A_{\rm gg,h}^{\rm S}(x) \otimes g^{(\nf}(x,m_h^2) \ ,
  \nn \\[0.3cm]
  (q_h+\bar{q}_{h})^{(n_f+1)}(x,m_h^2)  &=
  A_{\rm hq}^{\rm S}(x)\otimes \Sigma^{(\nf}(x,m_h^2) 
  + A_{\rm hg}^{\rm S}(x)\otimes g^{(\nf}(x,m_h^2)\ ,  \nn \\[0.3cm]
  \Sigma^{(n_f+1)}(x,m_h^2)  &= \Sigma^{\, (\nf}(x,m_h^2) + \left[ A^{\rm NS,+}_{qq,h}(x) + A^{\rm PS}_{qq,h}(x) + A_{\rm hq}^{\rm S}(x)\right] \otimes \Sigma^{(\nf}(x,m_h^2) \nn \\
  & + \left[ A^{\rm S}_{qg,h}(x) + A_{\rm hg}^{\rm S}(x) \right] \otimes g^{(\nf}(x,m_h^2)
\end{align}
with $q_h=\bar{q}_h$. Up to N$^3$LO the matching coefficients have the
following expansions in $\alpha_s$
\begin{align}
  A^{\rm NS,\pm}_{qq,h}(x) & = \lp \frac{\alpha_s(m_h^2)}{2\pi} \rp^2
  A^{\rm NS,\pm,(2)}_{qq,h}(x) + \lp \frac{\alpha_s(m_h^2)}{2\pi} \rp^3
  A^{\rm NS,\pm,(3)}_{qq,h}(x) \ , \nn \\
  A^{\rm S}_{gk,h}(x) & = \lp \frac{\alpha_s(m_h^2)}{2\pi} \rp^2
  A^{\rm S,(2)}_{gk,h}(x) + \lp \frac{\alpha_s(m_h^2)}{2\pi} \rp^3
  A^{\rm S,(3)}_{gk,h}(x), \quad k=q,g \ , \nn \\
  A^{\rm S}_{hk}(x) & = \lp \frac{\alpha_s(m_h^2)}{2\pi} \rp^2
  A^{\rm S,(2)}_{hk}(x) + \lp \frac{\alpha_s(m_h^2)}{2\pi} \rp^3
  A^{\rm S,(3)}_{hk}(x), \quad k=q,g \ , \nn \\
  A^{\rm PS}_{qq,h}(x) & = \lp \frac{\alpha_s(m_h^2)}{2\pi} \rp^3
  A^{\rm PS,(3)}_{qq,h}(x) \, , \nn \\
  A^{\rm S}_{qg,h}(x) & = \lp \frac{\alpha_s(m_h^2)}{2\pi} \rp^3
  A^{\rm S,(3)}_{qg,h}(x)\,.
\end{align}
At $\mathcal{O}(\alpha_S^2)$ we have that $A^{\rm NS,+}_{qq,h}(x) =
A^{\rm NS,-}_{qq,h}(x)$ whereas they start to differ at
$\mathcal{O}(\alpha_S^3)$. The NNLO matching coefficients were
computed in \cite{NNLO-MTM}\footnote{We thank the authors for the
code corresponding to the calculation.} and the N$^3$LO matching
coefficients
in~\cite{Bierenbaum:2009mv,Ablinger:2010ty,Kawamura:2012cr,Blumlein:2012vq,ABLINGER2014263,Ablinger:2014nga,Ablinger:2014vwa,Behring:2014eya,Ablinger:2019etw,Behring:2021asx,Fael:2022miw,Ablinger:2023ahe,Ablinger:2024xtt,BlumleinCode}.\footnote{We
thank Johannes Bl\"umlein for sharing a pre-release version the code from
Ref.~\cite{BlumleinCode} with us, which also contains code associated
with Refs.~\cite{Ablinger:2024xtt,Fael:2022miw}.
}
Notice that the above
conditions will lead to small discontinuities of the PDFs in its
evolution in $Q^2$, which are cancelled by similar matching terms in
the coefficient functions in massive VFN schemes, resulting in continuous physical
observables. In particular, the heavy-quark PDFs start from a non-zero
value at threshold at NNLO, which sometimes can even be negative.

The corresponding N$^3$LO relation for the matching of the $\MSbar$
coupling constant at the heavy quark threshold $m^2_h$ is given by 
\begin{equation}
\label{eq:as-nf1}
  \as^{\, (n_f+1)}(m_h^2) \: = \:
  \as^{\, (\nf} (m_h^2) +   C_2 \lp \frac{\as^{\, (\nf} (m_h^2)}{2\pi} \rp^3+   C_3 \lp \frac{\as^{\, (\nf} (m_h^2)}{2\pi} \rp^4
   \:\: ,
\end{equation}
where the matching coefficients $C_2$ and $C_3$ were computed in
\cite{Chetyrkin:1997sg,Chetyrkin:1997un}.
The value and the form of the matching coefficients in
eqs.~(\ref{eq:lp-nf1},\ref{eq:hp-nf1}) depend on the scheme used for
the quark masses; by default in \hoppet quark masses are taken to be
pole masses, though the option exists for the user to supply and have
thresholds crossed at $\MSbar$ masses, but only up to NNLO. We note
that in the current implementation in \hoppet the matching can only be
performed at the matching point that corresponds to the heavy-quark masses
themselves.

Both evolution and threshold matching preserve the momentum sum rule
\begin{equation}
  \int_0^1 dx~x \lp \Sigma(x,Q^2)+g(x,Q^2)\rp =1 \,,
\end{equation}
and valence sum rules
\begin{equation}
  \int_0^1 dx\, \left[q(x,Q^2)-{\bar q}(x,Q^2) \right] = \left\{ 
    \begin{array}{ll}
      1, & \text{for } q = d \text{ (in proton)}\\
      2, & \text{for } q = u \text{ (in proton)}\\
      0, & \text{other flavours}
    \end{array}
    \right.
\end{equation}
as long as they hold at the initial scale (occasionally not the case,
\eg in modified LO sets for Monte Carlo
generators~\cite{Sherstnev:2008dm}).

The default basis for the PDFs, called the \ttt{human} 
representation in \hoppet, is such that 
 the entries in an array
\ttt{pdf(-6:6)} of PDFs correspond to:
\bea 
\bar t={-6} \ ,  \bar b={-5} \ ,  \bar c={-4}
\ , \nn   \bar s&=&{-3} \ , \nn  \bar u={-2} \ , \nn
 \bar d={-1} \ , \\  g&=&{0} \ , \\ \nn   d={1} \ , \nn  u={2} 
\ , \nn  
s={3} \ , \nn   c&=&{4} \ , \nn b={5} \ , \nn  t={6} \ . \nn 
\eea
%
However, this representation leads
to a complicated form of the evolution equations.
The splitting matrix can be simplified considerably (made diagonal
except for a $2\times2$ singlet block) by switching to a different
flavour representation, which is named
the \ttt{evln} representation, for the PDF set, as explained in detail in
\cite{vanNeerven:1999ca,vanNeerven:2000uj}. This representation
is described in Table \ref{eq:diag_split}.

In the {\tt evln} basis, 
the gluon evolves coupled to the singlet  PDF $\Sigma$,
and all non-singlet PDFs evolve independently.
Notice that the representations of the PDFs
are preserved under linear operations, so in particular
they are preserved under DGLAP evolution.
The conversion from the \ttt{human} to the \ttt{evln}
representations of PDFs requires that the number of
active quark flavours $n_f$ be specified by the user, as described in
\ifreleasenote
Section~5.1.2 of Ref.~\cite{Salam:2008qg}.
\fi

\begin{table}
\begin{center}
\begin{tabular}{|r | c | l |}
\hline
     i & \mbox{name} & $q_i$ \\ \hline
     $ -6\ldots-(n_f+1)$ & $q_i$ & $q_i$\\
     $-n_f\ldots -2$ & $q_{\mathrm{NS},i}^{-}$ & 
$(q_i -  {\bar q}_i) - (q_1 - {\bar q}_1)$\\
      -1           & $q_{\mathrm{NS}}^{V}$ & 
$\sum_{j=1}^{n_f} (q_j -  {\bar q}_j)$\\
       0           & g & \textrm{gluon} \\
       1           & $\Sigma$ & $\sum_{j=1}^{n_f} (q_j +  {\bar q}_j)$\\
     $2\ldots n_f$ & $q_{\mathrm{NS},i}^{+}$ &
$ (q_i +  {\bar q}_i) - (q_1 + {\bar q}_1)$\\
      $(n_f+1)\ldots6$ & $q_i$ & $q_i$ \\
\hline
\end{tabular}
\caption{}{\label{eq:diag_split} The evolution representation 
(called \ttt{evln} in \hoppet)
of PDFs with $n_f$ active quark flavours
in terms of the \ttt{human} representation.}  
\end{center}
\end{table}

In \hoppet, unpolarised DGLAP evolution is available up to N$^3$LO
in the $\MSbar$ scheme, while for the DIS scheme
only evolution up to NLO is available, but without the NLO heavy-quark
threshold matching conditions. For polarised evolution up to NLO only
the $\MSbar$ scheme is available. The variable \ttt{factscheme}
takes different values for each factorisation scheme:
\begin{center}
  \begin{tabular}{|c|l|}\hline
    \ttt{factscheme} & Evolution\\[2pt]\hline
    1 & unpolarised $\MSbar$ scheme\\[2pt]\hline
    2 & unpolarised DIS scheme\\[2pt]\hline
    3 & polarised $\MSbar$ scheme\\\hline
  \end{tabular}
\end{center}
Note that mass thresholds are currently
missing in the DIS scheme.

The extension to QED is conceptually straightforward.
Further discussion of that is given in
Section~\ref{sec:qed-evolution}.

\section{Brief summary of \hoppet structure}
\label{sec:hoppet-structure}

\hoppet works in $x$-space.
It represents PDFs and splitting functions on grids, typically
multiple nested grids, each uniform in $y= \ln 1/x$, with the nesting
involving smaller spacings and smaller $y$ ranges so as to achieve
good accuracy not just at small $x$ but also large $x$.
The underlying convolutions of splitting functions with PDFs
effectively use piecewise polynomial interpolations of the PDFs.
The convolutions of the splitting functions with individual basis
polynomials are pre-evaluated using adaptive Gaussian integration.
Evolution equations are solved using Runge-Kutta methods.

\begin{table}[pt!]
  \centering
  \newcommand{\tabsecheader}[1]{\multicolumn{2}{c}{\bf #1\hspace{5.5em}\mbox{ }}}
\begin{tabular}{lp{0.55\textwidth}}
 \toprule
\tabsecheader{Configuration (optional)} \\\midrule
\begin{lstlisting}
hoppetSetExactDGLAP(threshold,split)
\end{lstlisting}
&\fn
Sets use of exact NNLO mass thresholds and splitting functions
  (default: both false).
  \\\hline
\begin{lstlisting}
hoppetSetApproximateDGLAPN3LO(variant)
hoppetSetSplittingN3LO(variant)
hoppetSetN3LOnfthresholds(variant)     
\end{lstlisting}
&\fn
  Sets variants for the choice of \ntlo splitting functions and
  thresholds, cf.\ Sec.~\ref{sec:n3lo-evolution}.
\\\hline
\begin{lstlisting}
hoppetSetQED(withqed,qcdqed,plq)
\end{lstlisting}
&  \fn Sets QED evolution and its options
  (cf.~Sec.~\ref{sec:qed-evolution}; default: all false).
\\
  \toprule
\tabsecheader{Initialisation} \\\midrule
\begin{lstlisting}
hoppetStart(dy,nloop)
\end{lstlisting} 
& \fn
Sets up a compound grid with
spacing in $\ln 1/x$ of \ttt{dy} at small $x$,
extending to $y = 12$ and numerical
order $\ttt=-6$. The $Q$ range for the tabulation will be $1\GeV <
Q<28 \TeV$, \fn~ \ttt{dlnlnQ=dy/4} and the factorisation scheme is
  ${\overline{\rm MS}}$ (\ttt{factscheme\_MSbar}).
  \\
\midrule
\begin{lstlisting}
hoppetStartExtended(ymax,dy,Qmin,
  Qmax,dlnlnQ,nloop,order,factscheme)
\end{lstlisting} & \fn
  More general initialisation. \\
\midrule 
\begin{lstlisting}
hoppetSetFFN(fixed_nf)
hoppetSetPoleMassVFN(mc,mb,mt)
hoppetSetMSbarMassVFN(mc,mb,mt)
\end{lstlisting} &
\fn Set heavy flavour scheme ($\MSbar$ available only to NNLO).\\
\toprule
\tabsecheader{Normal evolution} \\\midrule
\begin{lstlisting}
hoppetEvolve(asQ0,Q0alphas,
           nloop,muR_Q,LHAsub,Q0pdf)
\end{lstlisting} &
\fn PDF evolution: specifies the coupling \ttt{asQ0} at a 
scale \ttt{Q0alphas}, 
  the number of loops for  evol., \ttt{nloop},
  the ratio (\ttt{muR\_Q}) of ren. to fact. scales,
 the name of a subroutine \ttt{LHAsub(x,Q,f}) that fills \ttt{f(-6:6)},
 and the scale
\ttt{Q0pdf} at which one starts the PDF evolution.
 Note:  \ttt{LHAsub} is only called at scale \ttt{Q0pdf}
and in \CPP \ttt{f[iflv]} spans \ttt{iflv=0..12}.
\\
\toprule
\tabsecheader{Cached evolution} \\\midrule
\begin{lstlisting}
hoppetPreEvolve(asQ,Q0alphas, 
                nloop,muR_Q,Q0pdf)
\end{lstlisting} & \fn
 Preparation of a cached evolution.\\
\midrule
\begin{lstlisting}
hoppetCachedEvolve(LHAsub)
\end{lstlisting} &
\fn  Perform cached evolution with the initial condition
     at \ttt{Q0pdf} from a routine \ttt{LHAsub} 
with LHAPDF-like interface.
 Note: \ttt{LHAsub} only called at scale \ttt{Q0pdf}.\\
\toprule
\tabsecheader{Evaluation} \\\midrule
\begin{lstlisting}
hoppetEval(x,Q,f)
\end{lstlisting} &
\fn On return, \ttt{f(-6:6)} contains all flavours of the PDF set
                   (multiplied by $x$). In \CPP, the array
                   indices span 0 to 12. Increase upper bound by 5
                   with QED.\\
\midrule
\begin{lstlisting}
hoppetEvalSplit(x,Q,iloop,nf,pf)
\end{lstlisting} &
\fn On return, \ttt{pf(-6:6)} contains the (cached) convolution of the
                   \ttt{iloop} splitting function ($1=\text{LO}$) with the tabulated
                   PDF for the given \texttt{nf}. One can chain splitting
                   functions up to $\order{\as^4}$, e.g.\ \texttt{iloop=31} gives
                   $P_\text{NNLO}\otimes P_\text{LO} \otimes f$.\\
\midrule
\begin{lstlisting}
hoppetAlphaS(Q)
\end{lstlisting} &
\fn Returns the coupling at scale $Q$. \\  \midrule
\begin{lstlisting}
hoppetWriteLHAPDFGrid(basename,
                          pdf_index)
\end{lstlisting}
            & \fn
              Write an LHAPDF grid file to \ttt{basename\_nnnn.dat} where
              \ttt{nnnn} is \ttt{pdf\_index}; if
              \ttt{pdf\_index} is
              0, also write a template \ttt{basename.info} file.\\
\toprule
\tabsecheader{Cleanup (optional)} \\\midrule
\begin{lstlisting}
hoppetDeleteAll()
\end{lstlisting}
            & \fn
              Deletes all storage allocated by the streamlined interface
  \\\midrule
\end{tabular}

  \caption{Core methods of the streamlined interface in Fortran and \CPP.
    In Python, \ttt{hoppetStart(...)} is to be replaced with
    \ttt{hoppet.Start(...)}, and routines like \ttt{hoppet.Eval(...)}
    and \ttt{LHAsub(x,Q)} return
    \texttt{f} rather than taking \texttt{f} as an argument and
    filling it.
    \label{tab:streamlined-interface}
  }
\end{table}

The code has two interfaces.
For simple usage, it provides a so-called ``streamlined''
interface, giving high-level access to the functionality that is most
widely needed.
It is available from Fortran, \CPP and, as of v2, Python
(cf.\ Sec.~\ref{sec:pyinterface}).
Its main routines are listed in Table~\ref{tab:streamlined-interface}.
The functionality includes evolving to fill a PDF tabulation and then
accessing that PDF tabulation at given $x,Q$ points.
For faster tabulation of many distinct initial conditions, one can
pre-determine (cache) the evolution operators between the different
$Q$ scales at which the PDF is tabulated, and then repeatedly apply
that cached evolution.
The streamlined interface also provides access to convolutions of the
various orders of splitting functions with the tabulated PDF.

\begin{table}[pt!]
  \centering
\newcommand{\tabtt}[1]{\footnotesize\texttt{#1}}
\begin{tabular}{ll}
\toprule
\lstinline|type(grid_def :: grid| & \fn $x-$space grid definition \\ 
\midrule
\lstinline|real(dp), pointer :: gluon(:)| & \fn Holds a 
`grid quantity' (\eg gluon PDF) \\
\lstinline|real(dp), pointer :: PDFset(:,:)| & \fn
Grid representation of a (13-flavour) PDF set \\
\midrule
\lstinline|type(grid_conv)  :: Pgg| 
 & \fn Convolution operator ({\it i.e.} splitting function) \\ 
\lstinline|type(split_mat)  :: Pmat|
 & \fn Splitting matrix (with full flavour structure)\\ 
 \lstinline|type(mass_threshold_mat) :: MTM|
 & \fn Heavy quark mass-threshold matrix\\ 
\lstinline|type(dglap_holder) :: dglap_h| & \fn DGLAP holder (\ie all
splitting and mass-threshold matrices) \\
\midrule
\lstinline|type(running_coupling) :: coupling| & \fn Running coupling \\
\lstinline|type(evln_operator) :: evop| & \fn Evolution operator (linked
list of split \&  mass-threshold matrices)\\
\lstinline|type(pdf_table) :: table| & \fn PDF set tabulated in $x$ \& $Q$\\
\bottomrule\normalsize
\end{tabular}

  \caption{Core objects in the general interface.
    \label{tab:general-interface}
  }
\end{table}

For more advanced usage there is a ``general'' interface (sometimes
called the object-oriented interface, though it is only partially so).
It gives access to the various low-level objects that are useful in
DGLAP evolution, such as splitting functions, splitting matrices,
tabulations of PDFs, etc.
The main objects are listed in Table~\ref{tab:general-interface}.
Up to v2.1, that interface is accessible only in modern Fortran,
though in due course we expect to extend it to other languages.
We refer the reader to the original manual~\cite{Salam:2008qg} for an
extended discussion of the general interface.
For some uses, it can be convenient to initialise \hoppet with the
streamlined interface and then access the underlying objects in
Fortran from the
\repolink{src/streamlined_interface.f90}{streamlined\_interface}
module. 

Various examples are available with the two sets of interfaces, to be
found in the \masterlink{examples/} directory of the repository.
This document will focus mostly on the streamlined interface, though
in a places we will also discuss key additions to the general interface.


\section{QCD evolution at N$^3$LO}
\label{sec:n3lo-evolution}
In recent years significant progress in determining the perturbative
components needed for unpolarised QCD evolution at N$^3$LO has been
made (cf.\ \ref{sec:pqcd} for technical details on the evolution at
N$^3$LO).
This has now reached the stage that two PDF groups have released fits
at approximate N$^3$LO (aN$^3$LO)
accuracy~\cite{McGowan:2022nag,NNPDF:2024nan} along with their
combination in Ref.~\cite{Cridge:2024icl}.
Of the three contributions that are needed at full N$^3$LO accuracy
only two are fully known.
In particular the four-loop
$\beta$-function~\cite{vanRitbergen:1997va,Czakon:2004bu} and
associated mass thresholds~\cite{Chetyrkin:1997sg} have been known for
a very long time.
On the other hand the intricate calculations of the three-loop
matching relations needed for both the single and two mass VFNS were
only very recently
completed~\cite{Bierenbaum:2009mv,Ablinger:2010ty,Kawamura:2012cr,Blumlein:2012vq,ABLINGER2014263,Ablinger:2014nga,Ablinger:2014vwa,Behring:2014eya,Ablinger:2019etw,Behring:2021asx,Ablinger:2023ahe,Ablinger:2024xtt,Ablinger:2025awb}.
Finally the four-loop splitting functions entering the DGLAP equation
are currently not known exactly, except for certain $n_f$-dependent
terms~\cite{Gracey:1994nn,Davies:2016jie,Moch:2017uml,Gehrmann:2023cqm,Falcioni:2023tzp,Gehrmann:2023iah,Kniehl:2025ttz}.
However, enough Mellin-moments have been computed that together with
the exact pieces just mentioned and known small- and large-$x$
behaviours, approximate splitting functions suitable for phenomenology
can be reliably
determined~\cite{McGowan:2022nag,NNPDF:2024nan,Moch:2021qrk,Falcioni:2023luc,Falcioni:2023vqq,Moch:2023tdj,Falcioni:2024xyt,Falcioni:2024qpd}.
We have therefore incorporated the aforementioned pieces, using code
that is publicly available with those references, with the intention of
updating the splitting functions as they become more precisely known.

\subsection{Interface}

For the specific implementation in \hoppet version 2.0.0 we rely on
the approximations computed in
Refs.~\cite{Davies:2016jie,Moch:2017uml,Falcioni:2023luc,Falcioni:2023vqq,Moch:2023tdj,Falcioni:2024xyt,Falcioni:2024qpd} (FHMPRUVV),
the implementation of the three-loop (single-mass) VFNS coefficients as found in
Ref.~\cite{BlumleinCode}, which contains code associated with
Refs.~\cite{Ablinger:2024xtt,Fael:2022miw} and our own implementation
of the four-loop running coupling.\footnote{The implementation does
not allow for non-standard values of the QCD Casimir invariants beyond
three loops.}
Since the implementation here extends core features of \hoppet that
were already available in version 1.1.0, very little is needed on the
side of the user to invoke the evolution.
Most importantly all routines that take an \ttt{nloop} argument,
e.g.\ \ttt{hoppetStart}, and \ttt{hoppetEvolve} (or
\ttt{InitDglapHolder}, \ttt{InitRunningCoupling} in the general
interface)  now support \ttt{nloop = 4}.
The user also has a few choices they can make in terms of the
splitting functions and mass thresholds used.

Firstly, there have been successively improved approximations to the
splitting functions.
The user can control which series of approximation to use by making a
call to the subroutine \ttt{hoppetSetApproximateDGLAPN3LO(splitting\_approx)}.
At the time of writing, three options are available for the \ttt{splitting\_approx}:
\begin{itemize}
\item \ttt{n3lo\_splitting\_approximation\_up\_to\_2310\_05744}, with the
  approximations in papers up to and including Ref.~\cite{Moch:2023tdj};
\item \ttt{n3lo\_splitting\_approximation\_up\_to\_2404\_09701}, with the
  approximations in papers up to and including
  Ref.~\cite{Falcioni:2024xyt};
\item \ttt{n3lo\_splitting\_approximation\_up\_to\_2410\_08089}, the default value at the time of
  writing, with the approximations in papers up to and including
  Ref.~\cite{Falcioni:2024qpd}.
\end{itemize}
These constants, and the others discussed in this section, are defined
in the \repolink{src/dglap_choices.f90}{dglap\_choices} module in
Fortran.
In \CPP they are in the \ttt{hoppet} namespace, as defined in
\masterlink{src/hoppet.h}.

The approximate splitting functions come with an uncertainty band.
The user has a choice between the two extremities of the band and an
average of those extremities.
They can make the choice by calling \ttt{hoppetSetSplittingN3LO(variant)}, where
\ttt{variant} is one of
\begin{itemize}
\item \ttt{n3lo\_splitting\_Nfitav}: the average (the default),
\item \ttt{n3lo\_splitting\_Nfiterr1}: one of the two extremities,
\item \ttt{n3lo\_splitting\_Nfiterr2}: the other of the two.
  extremities
\end{itemize}

%
%

Besides the splitting functions, we have also incorporated the
single-mass thresholds up to N$^3$LO, as calculated in
Refs.~\cite{Bierenbaum:2009mv,Ablinger:2010ty,Kawamura:2012cr,Blumlein:2012vq,ABLINGER2014263,Ablinger:2014nga,Ablinger:2014vwa,Behring:2014eya,Ablinger:2019etw,Behring:2021asx,Ablinger:2023ahe,Ablinger:2024xtt}.
As of version 2.1.0, the \ntlo thresholds are available in two forms,
using code from Refs.~\cite{BlumleinCode,Fael:2022miw}.
One can choose which form to use by calling
\ttt{hoppetSetN3LOnfthresholds(n3lo\_threshold\_choice)},
with one of the following values for \ttt{n3lo\_threshold\_choice}:
\begin{itemize}
\item \ttt{n3lo\_nfthreshold\_libOME}:
  this uses piecewise high-order Laurent-series expansions in different
  regions of $x$, with a stated accuracy of at least $2048\epsilon$
  where $\epsilon$ is the precision of a 64-bit real variable.
  It relies on the
  \ttt{libome} \CPP library from \url{https://gitlab.com/libome/libome}.
  Initialisation time is below $1\,\text{s}$.
  This is the default choice.
  
\item \ttt{n3lo\_nfthreshold\_exact\_fortran}: all contributions are
  exact, except for the $A_\text{hg}^\text{S}$ term in
  Eq.~(\ref{eq:hp-nf1}), which is based on piecewise expansions.
  The exact contributions make extensive use of
  \texttt{hplog5}~\cite{FortranPolyLog} calls and lead to an
  initialisation time of $10{-}30\,\text{s}$.
  The piecewise expansions for $A_\text{hg}^\text{S}$ are less
  accurate than with the \ttt{n3lo\_nfthreshold\_libOME} choice.
\end{itemize}

%
%

\subsection{Reference results}

In Table~\ref{tab:n3lo-evolve} we show the results of the full N$^3$LO
evolution in the VFNS, using the most up-to-date perturbative input at
the time of writing this release note. To assess the evolution we take
the initial condition of Ref.~\cite{Dittmar:2005ed} at an initial
scale $\sqrt{2}~\text{GeV}$ and $n_f=3$. We evolve to
$Q=100~\text{GeV}$. The numbers are obtained using parametrised NNLO
splitting functions, but exact mass thresholds at this order.
The table has been generated with $\texttt{dy}=0.05$,
$\texttt{dlnlnQ}=\text{dy/4}$.
Increasing $\texttt{dy}=0.10$ leaves the results unchanged at the
precision shown, while going to $\texttt{dy}=0.20$ for higher speed
would change the results by a relative amount below $10^{-4}$.

Table~\ref{tab:n3lo-evolve} cannot be directly compared to the
benchmarking tables of Ref.~\cite{Cooper-Sarkar:2024crx} because they
do not include the mass thresholds in the \ntlo evolution.
To facilitate comparisons we therefore additionally provide
Table~\ref{tab:n3lo-evolve-nf4}, corresponding to fixed-flavour ($n_f=4$)
evolution, with the choice of
\ttt{n3lo\_splitting\_approximation\_up\_to\_2310\_05744}.
Of the MSHT and NNPDF results, the NNPDF results (Table~2 of
Ref.~\cite{Cooper-Sarkar:2024crx}) are closer to ours.

The results in both tables can be regenerated with the help
of the following script, which uses the Python interface of
Section~\ref{sec:pyinterface}:
\masterlink{benchmarking/tabulation_crosscheck_2406_16188.py}.

\begin{table}[p]
  \small \centering
  \begin{tabular}{c|cccccccc}
    $x$ &  $u-\bar u$ &$d-\bar d$ &$\bar d-\bar u$ &  $ 2(\bar u+\bar d)$  &$s+\bar s$&$c+\bar c$ &$b+\bar b$  &   $g$\\
    \toprule
$10^{-7}$ &  $1.0589^{-4}$ &  $4.8664^{-5}$ &  $8.1967^{-6}$ &  $1.6234^{+2}$ &  $8.0100^{+1}$ &  $7.7207^{+1}$ &  $6.5254^{+1}$ &  $1.1238^{+3}$ \\
$10^{-6}$ &  $5.9691^{-4}$ &  $3.2634^{-4}$ &  $3.3146^{-5}$ &  $7.6786^{+1}$ &  $3.7544^{+1}$ &  $3.5836^{+1}$ &  $2.9889^{+1}$ &  $5.1159^{+2}$ \\
$10^{-5}$ &  $3.0235^{-3}$ &  $1.7532^{-3}$ &  $1.3081^{-4}$ &  $3.5436^{+1}$ &  $1.7044^{+1}$ &  $1.6106^{+1}$ &  $1.3142^{+1}$ &  $2.2224^{+2}$ \\
$10^{-4}$ &  $1.4079^{-2}$ &  $8.2354^{-3}$ &  $4.9511^{-4}$ &  $1.5611^{+1}$ &  $7.2731^{+0}$ &  $6.7862^{+0}$ &  $5.3294^{+0}$ &  $8.8594^{+1}$ \\
$10^{-3}$ &  $6.0849^{-2}$ &  $3.5086^{-2}$ &  $1.7751^{-3}$ &  $6.3823^{+0}$ &  $2.7798^{+0}$ &  $2.5204^{+0}$ &  $1.8516^{+0}$ &  $3.0349^{+1}$ \\
$10^{-2}$ &  $2.3361^{-1}$ &  $1.3074^{-1}$ &  $5.8324^{-3}$ &  $2.2673^{+0}$ &  $8.5415^{-1}$ &  $7.0444^{-1}$ &  $4.6228^{-1}$ &  $7.7859^{+0}$ \\
$0.1$    &  $5.4846^{-1}$ &  $2.6950^{-1}$ &  $9.9965^{-3}$ &  $3.8453^{-1}$ &  $1.1248^{-1}$ &  $6.8296^{-2}$ &  $3.7899^{-2}$ &  $8.4964^{-1}$ \\
$0.3$    &  $3.4441^{-1}$ &  $1.2761^{-1}$ &  $2.9457^{-3}$ &  $3.4575^{-2}$ &  $8.8873^{-3}$ &  $3.9659^{-3}$ &  $2.0846^{-3}$ &  $7.8697^{-2}$ \\
$0.5$    &  $1.1790^{-1}$ &  $3.0597^{-2}$ &  $3.6526^{-4}$ &  $2.3206^{-3}$ &  $5.6808^{-4}$ &  $2.0185^{-4}$ &  $1.1382^{-4}$ &  $7.6337^{-3}$ \\
$0.7$    &  $1.9329^{-2}$ &  $2.9648^{-3}$ &  $1.2848^{-5}$ &  $5.2429^{-5}$ &  $1.2662^{-5}$ &  $3.4020^{-6}$ &  $2.4956^{-6}$ &  $3.7094^{-4}$ \\
$0.9$    &  $3.3153^{-4}$ &  $1.6737^{-5}$ &  $8.0961^{-9}$ &  $2.5214^{-8}$ &  $6.6432^{-9}$ &  $7.6153^{-10}$ &  $1.4323^{-9}$ &  $1.1716^{-6}$ 
  \end{tabular}
  \caption{N$^3$LO evolution of the initial condition given in Section
    4.4 of Ref.~\cite{Dittmar:2005ed}, using the same notation where
    $a\cdot10^{b} = a^b$. The evolution is performed taking the
    initial condition at $\sqrt{2}~\text{GeV}$ (just below the charm
    mass) and evolving in the VFNS up to $Q=100~\text{GeV}$, with a
    charm-quark pole mass of $1.414213563\GeV$ and a bottom-quark
    pole mass of $4.5\GeV$.
    The NNLO splitting functions are the parametrised form
    (\texttt{nnlo\_splitting\_variant = nnlo\_splitting\_param}), to 
    facilitate comparisons by other groups, and the \ntlo splitting
    functions use the
    \ttt{n3lo\_splitting\_approximation = n3lo\_splitting\_approximation\_up\_to\_2410\_08089} choice.}
  \label{tab:n3lo-evolve}
\end{table}

\begin{table}[p]
  \small \centering
  \begin{tabular}{c|cccccccc}
    $x$ &  $u-\bar u$ &$d-\bar d$ &$\bar d-\bar u$ &  $ 2(\bar u+\bar d)$  &$s-\bar s$ &$s+\bar s$&$c+\bar c$ &   $g$\\
    \toprule
$10^{-7}$ &  $9.8370^{-5}$ &  $4.5171^{-5}$ &  $7.5013^{-6}$ &  $1.4888^{+2}$ &  $-2.9105^{-5}$ &             $7.3368^{+1}$ &  $7.2653^{+1}$ &  $1.0851^{+3}$ \\
$10^{-6}$ &  $5.6405^{-4}$ &  $3.0895^{-4}$ &  $3.0730^{-5}$ &  $7.1925^{+1}$ &  $-4.6739^{-5}$ &             $3.5111^{+1}$ &  $3.4544^{+1}$ &  $5.0392^{+2}$ \\
$10^{-5}$ &  $2.8946^{-3}$ &  $1.6810^{-3}$ &  $1.2302^{-4}$ &  $3.3868^{+1}$ &  $-3.5766^{-6}$ &             $1.6258^{+1}$ &  $1.5808^{+1}$ &  $2.2292^{+2}$ \\
$10^{-4}$ &  $1.3633^{-2}$ &  $7.9832^{-3}$ &  $4.7274^{-4}$ &  $1.5188^{+1}$ &  $\phantom{-}2.1123^{-4}$ &   $7.0599^{+0}$ &  $6.7033^{+0}$ &  $9.0268^{+1}$ \\
$10^{-3}$ &  $5.9567^{-2}$ &  $3.4382^{-2}$ &  $1.7232^{-3}$ &  $6.3028^{+0}$ &  $\phantom{-}3.9314^{-4}$ &   $2.7387^{+0}$ &  $2.4621^{+0}$ &  $3.1350^{+1}$ \\
$10^{-2}$ &  $2.3130^{-1}$ &  $1.2962^{-1}$ &  $5.7645^{-3}$ &  $2.2675^{+0}$ &  $-1.9644^{-4}$ &             $8.5255^{-1}$ &  $6.6402^{-1}$ &  $8.1568^{+0}$ \\
$0.1$    &  $5.5131^{-1}$ &  $2.7140^{-1}$ &  $1.0085^{-2}$ &  $3.8980^{-1}$ &  $-3.1812^{-4}$ &             $1.1388^{-1}$ &  $5.9843^{-2}$ &  $9.0615^{-1}$ \\
$0.3$    &  $3.5044^{-1}$ &  $1.3015^{-1}$ &  $3.0145^{-3}$ &  $3.5426^{-2}$ &  $-3.8409^{-5}$ &             $9.0900^{-3}$ &  $3.3507^{-3}$ &  $8.4431^{-2}$ \\
$0.5$    &  $1.2112^{-1}$ &  $3.1518^{-2}$ &  $3.7779^{-4}$ &  $2.3973^{-3}$ &  $-3.3053^{-6}$ &             $5.8501^{-4}$ &  $1.7709^{-4}$ &  $8.1568^{-3}$ \\
$0.7$    &  $2.0078^{-2}$ &  $3.0889^{-3}$ &  $1.3448^{-5}$ &  $5.4598^{-5}$ &  $-1.1810^{-8}$ &             $1.3105^{-5}$ &  $3.6963^{-6}$ &  $3.9248^{-4}$ \\
$0.9$    &  $3.5128^{-4}$ &  $1.7793^{-5}$ &  $8.6472^{-9}$ &  $2.6378^{-8}$ &  $-1.5848^{-10}$ &            $6.8222^{-9}$ &  $2.6776^{-9}$ &  $1.2262^{-6}$ 
  \end{tabular}
  \caption{N$^3$LO evolution of the initial condition given in Section
    4.4 of Ref.~\cite{Dittmar:2005ed}, using the same notation where
    $a\cdot10^{b} = a^b$. The evolution is performed taking the
    initial condition at $\sqrt{2}~\text{GeV}$ (just below the charm
    mass) and evolving in the FFN scheme ($n_f = 4$) up to
    $Q=100~\text{GeV}$.
    The NNLO splitting functions are the parametrised form
    (\texttt{nnlo\_splitting\_variant = nnlo\_splitting\_param}), to 
    facilitate comparisons by other groups, and the \ntlo splitting
    functions use the
    \ttt{n3lo\_splitting\_approximation =
      n3lo\_splitting\_approximation\_up\_to\_2310\_05744} choice.
    This table can be directly compared to tables 1 and 2 in Ref.~\cite{Cooper-Sarkar:2024crx}}
  \label{tab:n3lo-evolve-nf4}
\end{table}


\begin{figure}[p]
  \centering
  \begin{subfigure}{0.49\textwidth}
    \centering
    \includegraphics[width=\textwidth,page=1]{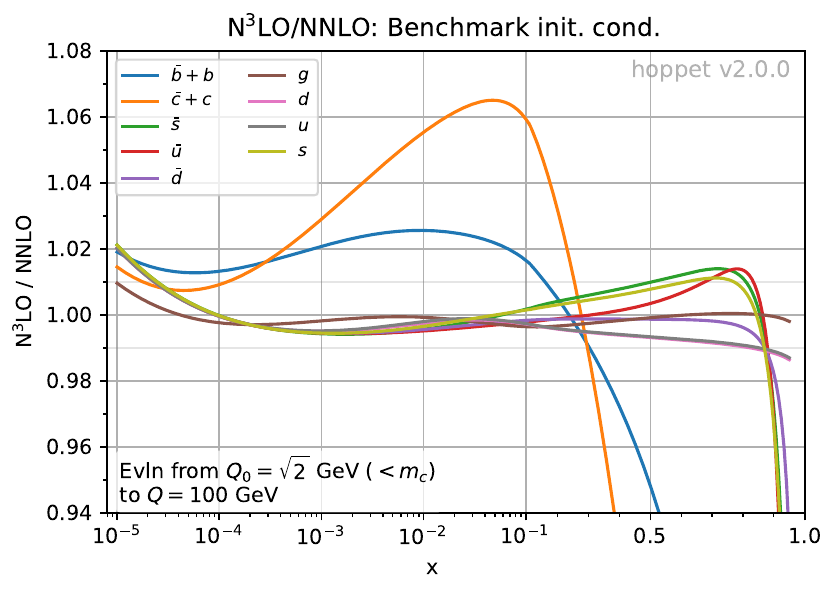}\\[-1.5ex]
    \caption{}
    \label{fig:n3lok-benchmark-mtm}
  \end{subfigure}
  \begin{subfigure}{0.49\textwidth}
    \centering
    \includegraphics[width=\textwidth,page=3]{figs-v2/n3lo_k_factors.pdf}\\[-1.5ex]
    \caption{}
    \label{fig:n3lok-benchmark-nomtm}
  \end{subfigure}\\[1ex]
  \begin{subfigure}{0.49\textwidth}
    \centering
    \includegraphics[width=\textwidth,page=4]{figs-v2/n3lo_k_factors.pdf}\\[-1.5ex]
    \caption{}
    \label{fig:n3lok-nnpdf40-mtm}
  \end{subfigure}
  \begin{subfigure}{0.49\textwidth}
    \centering
    \includegraphics[width=\textwidth,page=6]{figs-v2/n3lo_k_factors.pdf}\\[-1.5ex]
    \caption{}
    \label{fig:n3lok-nnpdf40-nomtm}
  \end{subfigure}
  \caption{Ratio of the PDFs at $100\GeV$ with \ntlo evolution versus
    NNLO evolution.
    The evolution starts from the same initial condition
    at NNLO and \ntlo, at an initial scale
    $Q_0\simeq 1.41\GeV$.
    In the upper plots, we use the standard benchmark initial
    condition.
    In the lower plots, the initial condition is the
    \ttt{NNPDF40\_pch\_nnlo\_as\_01180} PDF~\cite{NNPDF:2021njg} at $Q_0$.
    The left-hand plots show the ratio with \ntlo evolution
    including \ntlo mass thresholds, while the right-hand plots show
    the evolution without the \ntlo mass thresholds.
    In all cases, the \ntlo evolution uses
    \ttt{n3lo\_splitting\_approximation\_up\_to\_2410\_08089}.  }
  \label{fig:n3lo-v-nnlo-ev}
\end{figure}

We close this section by illustrating the impact of \ntlo versus NNLO
evolution, Fig.~\ref{fig:n3lo-v-nnlo-ev}.
Let us first focus on the top-left plot,
Fig.~\ref{fig:n3lok-benchmark-mtm}.
We start with the benchmark initial condition at the standard low
scale of $Q_0 = \sqrt{2}\GeV$, just below a charm mass of $m_c =
1.414213563 \GeV$, so as to ensure that the evolution starts with $n_f
= 3$.
We then evolve the PDF separately with \ntlo and NNLO evolution, and
show the \ntlo/NNLO ratio at $Q = 100\GeV$.
Each line corresponds to a different flavour.
For light flavours, in the range that is relevant to the LHC at
central rapidities, $10^{-4} \lesssim x \lesssim 0.5$, the effect of
\ntlo corrections on the evolution of the light-flavour PDFs is
generally below a percent, and typically less than or around half a
percent.
For heavy flavour, the effect is much more significant, with a
$\sim 6\%$ effect on the charm distribution for $0.01 \lesssim x
\lesssim 0.1$ and about $-15\%$ at $x=0.5$.

Fig.~\ref{fig:n3lok-benchmark-nomtm} is the analogous plot with the
\ntlo mass-threshold contributions turned off.
It illustrates that the large effects on the charm and bottom PDFs are
a consequence mainly of \ntlo mass-thresholds, not the \ntlo splitting functions.
Comparing Figs.~\ref{fig:n3lok-benchmark-mtm} and
\ref{fig:n3lok-benchmark-nomtm} for light flavours, one sees that the $0.5\%$
effects are coming both from the \ntlo splitting functions and the
\ntlo mass thresholds.

Finally, Figs.~\ref{fig:n3lok-nnpdf40-mtm} and
\ref{fig:n3lok-nnpdf40-nomtm} show analogous plots with an initial
condition taken from the \ttt{NNPDF40\_pch\_nnlo\_as\_01180} PDF
set~\cite{NNPDF:2021njg} at a similar $Q_0 = 1.41 \GeV$ (again below
the charm threshold, $m_c = 1.51\GeV$).
The results are broadly similar, showing that our conclusions about the
size of \ntlo effects are robust with respect to the choice of PDF.

\section{Hadronic Structure Functions}
\label{sec:structure-funcs}
As of \hoppet version 2.0.0, the code provides access to the massless
hadronic structure functions. The structure functions are expressed as
convolutions of a set of massless hard coefficient functions and PDFs,
and make use of the tabulated PDFs and streamlined interface.  They
are provided such that they can be used directly for cross section
computations in DIS or VBF, as implemented for example in {\tt
  disorder}~\cite{Karlberg:2024hnl} and the {\tt proVBFH}
package~\cite{Cacciari:2015jma,Dreyer:2016oyx,Dreyer:2018qbw,Dreyer:2018rfu,Dreyer:2020urf,Dreyer:2020xaj}.

The massless structure functions have been found to be in good
agreement with those that can be obtained with
\APFELPP{}~\cite{Bertone:2013vaa,Bertone:2017gds} (at the level of
$10^{-5}$ relative precision).
The benchmarks with \APFELPP{} and the code used to carry them out are
described in detail in Ref.~\cite{Bertone:2024dpm} and at
\url{https://github.com/alexanderkarlberg/n3lo-structure-function-benchmarks}. A
simplified version of that benchmark is also included in the \hoppet
repository and can be found in
\masterlink{benchmarking/structure\_functions\_benchmark\_checks.f90}.
Technical details on the implementation of the structure functions in
\hoppet can be found in
Refs.~\cite{Dreyer:2016vbc,Karlberg:2016zik,Bertone:2024dpm}, and here
we mainly focus on the code interface.

The structure functions have been implemented including only QCD
corrections up to N$^3$LO using both the exact and parametrised
coefficient functions found in
Refs.~\cite{vanNeerven:1999ca,vanNeerven:2000uj,Moch:2004xu,Vermaseren:2005qc,Moch:2008fj,Davies:2016ruz,Blumlein:2022gpp},\footnote{Note
that the piece presented in Ref.~\cite{Davies:2016ruz} has not been
given in exact form. Only a parametrised version is available and
is what is being used in \hoppet.}  and can make use of PDFs evolved
at N$^3$LO as described in Sec.~\ref{sec:n3lo-evolution}. The
structure functions can also be computed using PDFs interfaced through
LHAPDF~\cite{LHAPDF}.

\subsection{Initialisation}
\label{sec:structure-funcs-init}

The structure functions can be accessed by using the
\repolink{src/structure_functions.f90}{structure\_functions} module.
They can also be accessed through the
streamlined interface by prefixing \ttt{hoppet}, as described
later in section~\ref{sec:structure-functions-streamlined}.
The description here corresponds to an intermediate-level interface, which
relies on elements such as the \ttt{grid} and splitting functions
having been initialised in the streamlined interface, through a call
to \ttt{hoppetStart} or \ttt{hoppetStartExtended}, cf.\
\ifreleasenote
Section~8 of Ref.~\cite{Salam:2008qg}.\footnote{Users needing a lower level
  interface should inspect the code in \masterlink{src/structure\_functions.f90}.}
\fi
After this initialisation has been carried out, one calls
\begin{lstlisting}
  call StartStrFct(order_max [, nflav] [, scale_choice] &
                  & [, constant_mu] [ ,param_coefs] [ ,wmass] [ ,zmass])
\end{lstlisting}
specifying as a minimum the perturbative order --- currently
\ttt{order\_max} $ \le 4$ (\ttt{order\_max} $ =1$ corresponds to LO).

If \ttt{nflav} is not passed as an argument, the structure functions
are initialised to support a variable flavour-number scheme (the
masses that are used at any given stage will be those set in the
streamlined interface).
Otherwise a fixed number of light flavours is used, as indicated by
\ttt{nflav}, which speeds up initialisation.
Note that specifying a variable flavour-number scheme only has an
impact on the evolution and on $n_f$ terms in the coefficient
functions.
The latter, however always assume massless quarks.
Hence in both the fixed and variable flavour-number scheme the
structure functions should not be considered phenomenologically
reliable if $Q$ is comparable to the quark mass.
%
Together \ttt{xR}, \ttt{xF}, \ttt{scale\_choice}, and \ttt{constant\_mu} control
the renormalisation and factorisation scales and the degree of
flexibility that will be available in choosing them at later stages.
Specifically the (integer) \ttt{scale\_choice}
argument should be one of the following values (defined in the
\ttt{structure\_functions} module):
\begin{itemize}
\item \ttt{scale\_choice\_Q} (default) means that the code will always
  use $Q$ multiplied by \ttt{xR} or \ttt{xF} as the renormalisation
  and factorisation scale respectively (with \ttt{xR} or \ttt{xF} as
  set at initialisation).
\item \ttt{scale\_choice\_fixed} corresponds to a fixed scale
  \ttt{constant\_mu}, multiplied by \ttt{xR} or \ttt{xF} as set at
  initialisation.
\item \ttt{scale\_choice\_arbitrary} allows the user to choose
  arbitrary scales at the moment of evaluating the structure
  functions.
  In this last case, the structure functions are saved as separate arrays, 
  one for each perturbative order, and with dedicated additional arrays
  for terms proportional to logarithms of $Q/\mu_F$. 
  This makes for a slower evaluation compared to the two other
  scale choices.
\end{itemize}
%

If \ttt{param\_coefs} is set to \ttt{.true.}\ (its default) then the
structure functions are computed using the NNLO and N$^3$LO
parametrisations found in
Refs.~\cite{vanNeerven:1999ca,vanNeerven:2000uj,Moch:2004xu,Vermaseren:2005qc,Moch:2008fj,Davies:2016ruz},
which are stated to have a relative precision of a few permille (order
by order) except at particularly small or large values of $x$.
Alternatively, it is possible to ask for the exact coefficient
functions.\footnote{The LO and NLO
coefficient functions are always exact as their expressions are very
compact. Note that for very large values of $x$ we switch to a
large-$x$ expansion for the regular part of non-singlet coefficient
functions at N$^3$LO, to avoid numerical instabilities in the exact
expressions as discussed in Appendix A of
Ref.~\cite{Bertone:2024dpm}.}
Since the expressions are large, and slow down the compilation, one
must explicitly request their compilation with the
\texttt{-DHOPPET\_USE\_EXACT\_COEF=ON} CMake flag.
Having done that, one then has the option of setting
\ttt{param\_coefs} to \ttt{.false.}.
This will lead the initialisation to become quite slow (up to two
minutes rather than a few seconds).
Given the good accuracy of the parametrised coefficient functions,
they are to be preferred for most applications.
%

The masses of the electroweak vector bosons are used only to calculate
the weak mixing angle, $\sin^2 \theta_W = 1 - (m_W/m_Z)^2$, which
enters in the neutral-current structure functions.

At this point all the tables that are needed for the structure
functions have been allocated.
In order to fill the tables, one first needs to set up the running
coupling and evolve the initial PDF with \ttt{hoppetEvolve}, as
described in
\ifreleasenote
Section~8.2 of Ref.~\cite{Salam:2008qg}.
\fi

%
%
With the PDF table filled in the streamlined interface one calls
\begin{lstlisting}
  call InitStrFct(order[, separate_orders] [, xR] [, xF] [, flavour_decomposition])
\end{lstlisting}
specifying the order at which one would like to compute the structure
functions.
The logical flag \ttt{separate\_orders} should be set to \ttt{.true.}\ if one
wants access to the individual coefficients of the perturbative
expansion as well as the sum up to some maximum order, \ttt{order}.
With \ttt{scale\_choice\_Q} and \ttt{scale\_choice\_fixed}, the
default of \ttt{.false.}\ causes only the sum over perturbative orders
to be stored.
This gives faster evaluations of structure functions because it is
only necessary to interpolate the sum over orders, rather than
interpolate one table for each order.
With \texttt{scale\_choice\_arbitrary}, the default is \ttt{.true.},
which is the only allowed option, because separate tables for each
order are required for the underlying calculations.

Finally, the optional flag \ttt{flavour\_decomposition} controls an
experimental feature of giving access to the structure functions
decomposed into their underlying quark flavours without the associated
vector boson couplings. It is currently only possible to access the
structure functions in this way up to NLO, and since the feature is
not fully mature we invite interested readers to inspect the source
code directly for more information.

\subsection{Accessing the Structure Functions}
\label{sec:structure-funcs-access}
At this point the structure functions can be accessed as in the following example
\begin{lstlisting}
  real(dp) :: ff(-6:7), x, Q, muR, muF
  
  call StartStrFct(order_max = 4, scale_choice = scale_choice_Q)
  [...]
  call InitStrFct(order_max = 4)
  ff = StrFct(x, Q[, muR] [, muF])
\end{lstlisting}
at the value $x$ and $Q$.
%
With \ttt{scale\_choice\_arbitrary}, the \ttt{muR} and \ttt{muF}
arguments must be provided.
With other scale choices, they do not need to be provided, but if they
are then they should be consistent with the original scale choice.
The structure functions in this example are stored in the
array \ttt{ff}. The components of this array can be accessed through
the indices 
\begin{lstlisting}
  integer, parameter :: iF1Wp = 1 !< F1 W+
  integer, parameter :: iF2Wp = 2 !< F2 W+
  integer, parameter :: iF3Wp = 3 !< F3 W+
  integer, parameter :: iF1Wm =-1 !< F1 W-
  integer, parameter :: iF2Wm =-2 !< F2 W-
  integer, parameter :: iF3Wm =-3 !< F3 W-
  integer, parameter :: iF1Z  = 4 !< F1 Z 
  integer, parameter :: iF2Z  = 5 !< F2 Z
  integer, parameter :: iF3Z  = 6 !< F3 Z
  integer, parameter :: iF1EM =-4 !< F1 $\gamma$
  integer, parameter :: iF2EM =-5 !< F2 $\gamma$
  integer, parameter :: iF1gZ = 0 !< F1 $\gamma$Z interference
  integer, parameter :: iF2gZ =-6 !< F2 $\gamma$Z interference
  integer, parameter :: iF3gZ = 7 !< F3 $\gamma$Z interference
\end{lstlisting}
For instance one would access the electromagnetic $F_1$ structure
function through \ttt{ff(iF1EM)}. It is returned at the \ttt{order\_max}
that was specified in \ttt{InitStrFct}.
The structure functions can also be accessed order by order if the
\ttt{separate\_orders} flag was set to \ttt{.true.} when initialising.
They are then obtained as follows
\begin{lstlisting}
  real(dp) :: flo(-6:7), fnlo(-6:7), fnnlo(-6:7), fn3lo(-6:7), x, Q, muR, muF
  [...]
  call InitStrFct(4, .true.)
  flo   = F_LO(x, Q, muR, muF)
  fnlo  = F_NLO(x, Q, muR, muF)
  fnnlo = F_NNLO(x, Q, muR, muF)
  fn3lo = F_N3LO(x, Q, muR, muF)
\end{lstlisting}
The functions return the individual contributions at each order in
$\as$, including the relevant factor of $\as^n$.
Hence the sum of \ttt{flo}, \ttt{fnlo}, \ttt{fnnlo}, and
\ttt{fn3lo} would be equal to the full structure function at N$^3$LO as
contained in \ttt{ff} in the example above.
Note that in the \ttt{F\_LO} etc.\ calls, the \ttt{muR} and \ttt{muF}
arguments are not optional and that when a prior scale choice has been
made (e.g. \ttt{scale\_choice\_Q}) they are required to be consistent
with that prior scale choice.

An example of structure function evaluations using the Fortran~90
interface is to be found in
\repolink{examples/f90/structure_functions_example.f90}{examples/f90/structure\_functions\_example.f90}. 

\subsection{Streamlined interface}
\label{sec:structure-functions-streamlined}
The structure functions can also be accessed through the streamlined
interface, so that they may be called for instance from C/\CPP{}. The
functions to be called are very similar to those described above.
For simple usage one can call 
\begin{lstlisting}
  call hoppetStartStrFct(order_max)
\end{lstlisting}
where \ttt{order\_max-1}
is the maximal
power of $\as$.
Alternatively, the extended version of the interface,
\ttt{hoppetStartStrFctExtended},  takes all the same arguments as
\ttt{StartStrFct} described above. One difference is that in order to
use a variable flavour scheme the user should set \ttt{nflav} to a
negative value. After evolving or reading in a PDF, the user then calls
\begin{lstlisting}
  call hoppetInitStrFct(order, separate_orders)
\end{lstlisting}
to initialise the actual structure functions. The structure functions
can then be accessed through the subroutines
\begin{lstlisting}
  real(dp) :: ff(-6:7), flo(-6:7), fnlo(-6:7), fnnlo(-6:7), fn3lo(-6:7), x, Q, muR, muF
  [...]
  call hoppetStrFct(x, Q, muR, muF, ff)        ! Full structure function
  call hoppetStrFctNoMu(x, Q, ff)              ! Full structure function, muR=muF=Q
  call hoppetStrFctLO(x, Q, muR, muF, flo)     ! LO term
  call hoppetStrFctNLO(x, Q, muR, muF, fnlo)   ! NLO term
  call hoppetStrFctNNLO(x, Q, muR, muF, fnnlo) ! NNLO term
  call hoppetStrFctN3LO(x, Q, muR, muF, fn3lo) ! N3LO term
\end{lstlisting}
The \CPP{} header contains indices for the structure functions and scale
choices, which are all in the \ttt{hoppet} namespace.
\begin{lstlisting}
  const int iF1Wp = 1+6;
  const int iF2Wp = 2+6;
  const int iF3Wp = 3+6;
  const int iF1Wm =-1+6;
  const int iF2Wm =-2+6;
  const int iF3Wm =-3+6;
  const int iF1Z  = 4+6;
  const int iF2Z  = 5+6;
  const int iF3Z  = 6+6;
  const int iF1EM =-4+6;
  const int iF2EM =-5+6;
  const int iF1gZ = 0+6;
  const int iF2gZ =-6+6;
  const int iF3gZ = 7+6;

  const int scale_choice_fixed     = 0;
  const int scale_choice_Q         = 1;
  const int scale_choice_arbitrary = 2;
\end{lstlisting}
Note that in \CPP{} the structure function indices start from 0 and that the \CPP{}
array that is to be passed to functions such as \ttt{hoppetStrFct}
would be defined as \ttt{double ff[14]}.

An example of structure function evaluations using the \CPP{} version
of the streamlined interface is to be found in
\repolink{examples/cpp/structure_functions_example.cc}{examples/cpp/structure\_functions\_example.cc}
and a Python example is similarly to be found at
\masterlink{examples/python/structure_function_example.py}.

\begin{figure}[tb!]
  \centering\includegraphics[width=0.49\textwidth]{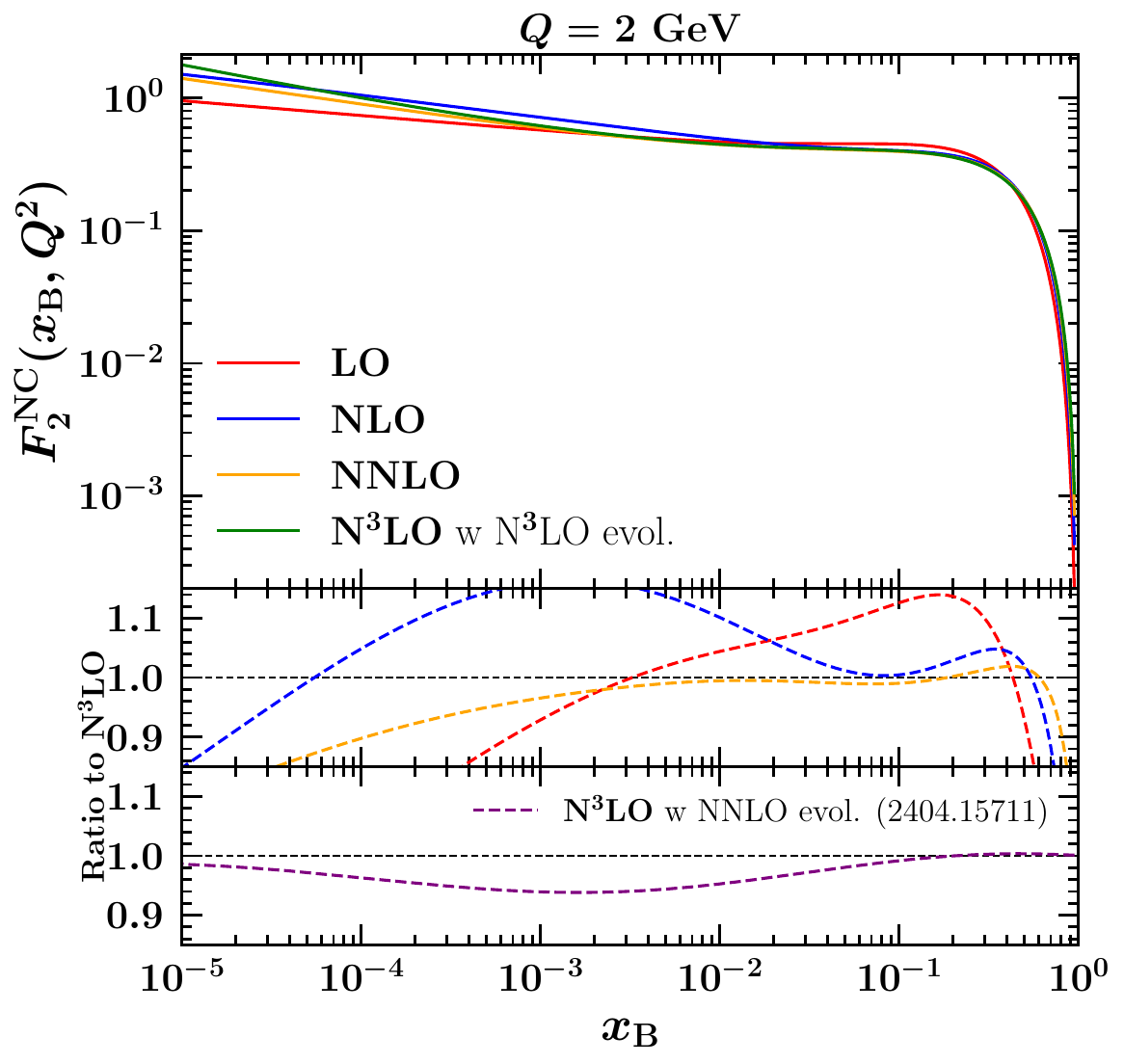}
  \centering\includegraphics[width=0.49\textwidth]{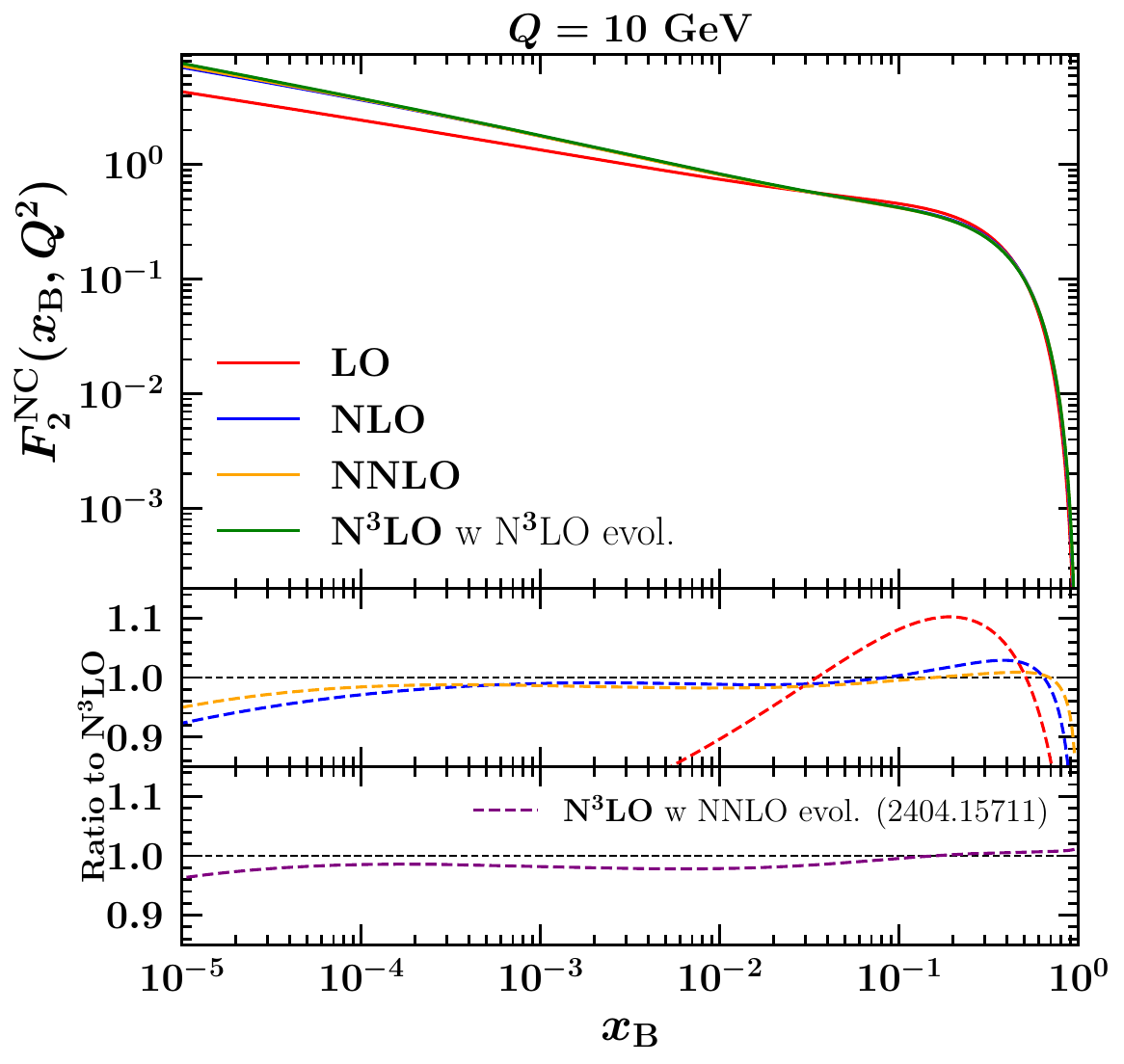}
  \caption{The structure function $F_2^{\rm NC}$ plotted as a function
    of $x_{\rm B}$ in the range $[10^{-5}:0.9]$ at $Q=2$~GeV (left)
    and $Q=10$~GeV (right). Each plot displays the curves at LO, NLO,
    NNLO, and N$^3$LO with the middle panel showing the ratio to
    N$^3$LO. The lowest panel shows the ratio of the \ntlo with NNLO
    evolution to the full \ntlo result. Adapted from
    Ref.~\cite{Bertone:2024dpm} using the full \ntlo evolution.}
  \label{fig:perturbativeconvergence}
\end{figure}

Finally we present a small update on the results presented in
Ref.~\cite{Bertone:2024dpm}. That reference was published before the
full \ntlo evolution had been implemented in \hoppet, and the \ntlo
structure functions were therefore obtained with NNLO evolution. A
full update of the tables and plots is beyond the scope of the current
work, but for illustrative purposes we present here an updated version
of Fig.~3 of Ref.~\cite{Bertone:2024dpm}.
The update uses the full \ntlo evolution and is shown as
Fig.~\ref{fig:perturbativeconvergence}.
It shows the full charged-lepton neutral current $F_2$ structure
function ($F_2^{\rm NC}$) at various perturbative orders, for two
values of $Q$ ($2$ GeV and $10$ GeV). It also shows the relative
difference of the various orders with respect to \ntlo, and in the
lower panel the ratio of the \ntlo result of
Ref.~\cite{Bertone:2024dpm} to that obtained here with the full
evolution.
As can be seen by comparing to the original figures and from the lowest
panel, the \ntlo evolution has the effect at low $Q$ of increasing the
\ntlo $F_2^{\rm NC}$ by a few percent, at least for $x$-values in the range
$10^{-4}$ to $10^{-1}$.
%

\section{Evolution including QED contributions}
\label{sec:qed-evolution}

The combined QED+QCD evolution, as implemented in \hoppet since
version 2.0.0 (and earlier in a dedicated \ttt{qed} branch), was first
described in
Refs.~\cite{Manohar:2016nzj,Manohar:2017eqh,Buonocore:2020nai,Buonocore:2021bsf}.
The determination of which contributions to include follows a
consistent approach based on the so-called ``phenomenological'' counting
scheme.
Within this scheme, one considers the QED coupling $\alpha$ to be of order $\as^2$, and takes the
photon (lepton) PDF to be of order $\alpha L$ ($\alpha^2 L^2$), where
$L$ is the logarithm of the
ratio of the factorisation scale to a typical hadronic scale and is considered to be of order $L \sim 1/\as$.
In contrast, quark and gluon
PDFs are considered to be of order $(\as L)^n={\cal O}(1)$.\footnote{The
  above counting is to be contrasted with a ``democratic'' scheme, in
  which one considers $\alpha \sim \as \sim 1/L$ and where the aim would be to maintain
  the same loop order across all couplings and splitting functions,
  regardless of the relative numbers of QCD and QED couplings that
  they involved.}
From this point of view, NNLO (3-loop) QCD evolution provides control of
terms of order up to $\as^{n+2} L^n \sim \as^2$.
To achieve a corresponding accuracy when including QED contributions,
\hoppet has been extended to account for
\begin{enumerate}
\item \label{item:qed1} 1-loop QED splitting functions~\cite{Roth:2004ti}, which first
  contribute at order $\alpha L \sim \as$, i.e.\ count as NLO QCD
  corrections;
  
\item \label{item:qed2} 1-loop QED running coupling, including lepton and quark
  thresholds, which first contributes at order $\alpha^2 L^2 \sim
  \as^2$, i.e. like NNLO QCD; 
  
\item \label{item:qed3} 2-loop mixed QCD-QED splitting
  functions~\cite{deFlorian:2015ujt},
  which first contribute at order
  $\alpha \as L \sim \as^2$, i.e.\ count as NNLO QCD corrections;

\item \label{item:qed4} optionally, the 2-loop pure QED $P_{\ell q}$ splitting
  function~\cite{deFlorian:2016gvk}, which brings absolute accuracy
  $\alpha^2 L\sim \as^3$ to the lepton distribution (which starts at
  $\alpha^2 L^2 \sim \as^2$).\footnote{The implementation of the 2-loop $P_{\ell q}$ splitting
  function in the hoppet code (\ttt{Plq\_02} in the code) was carried out by Luca Buonocore.}
  In an absolute counting of accuracy, this is not needed.
  However, if one wants lepton distributions to have the same relative
  NLO accuracy as the photon distribution, it should be
  included.
\end{enumerate}
The code could be extended systematically to aim at a higher
accuracy. For instance, if one wished to reach N$^3$LO accuracy in the
phenomenological counting, one would need to include 3-loop mixed
QCD-QED splitting functions at order $\alpha \as^2$, which contribute
at order $\alpha \as^2 L \sim \as^3$ but are currently not
available, the full 2-loop pure QED splitting functions~\cite{deFlorian:2016gvk}, 
and the 2-loop mixed QED-QCD contributions to the running
of the QED coupling, which contribute at order $\alpha^2 \as L^2 \sim \as^3$
(see e.g.\ \cite{Cieri:2018sfk}). 
%
%

The rest of this section is structured as follows:
Section~\ref{sec:qed-streamlined} shows how to get QED evolution in
the streamlined interface, which is relatively straightforward.
Section~\ref{sec:qed-implementation} then gives a technical discussion
of the implementation of the QED evolution, including details, for
example, regarding the choice of QED coupling.
Some readers may prefer to skip or skim this on a first reading.

\subsection{Streamlined interface with QED effects} 
\label{sec:qed-streamlined}
The streamlined interface including QED effects works as in the case of pure QCD evolution.
One has to add the following call
\begin{lstlisting}
  logical use_qed, use_qcd_qed, use_Plq_nnlo
  ...
  call hoppetSetQED(use_qed, use_qcd_qed, use_Plq_nnlo)
\end{lstlisting}
before using the streamlined interface routines.
The \ttt{use\_qed} argument turns QED evolution on/off at order
${\cal O}(\alpha)$ (i.e.\ items \ref{item:qed1} and \ref{item:qed2} in
the enumerated list at the beginning of
Sec.~\ref{sec:qed-evolution}).
The \ttt{use\_qcd\_qed} one turns
mixed QCD$\times$QED effects on/off in the evolution (i.e.\ item \ref{item:qed3}) and \ttt{use\_Plq\_nnlo} turns 
the order $\alpha^2 P_{\ell q}$ splitting function on/off (i.e.\
item \ref{item:qed4}).
Without this call, all QED corrections
are off.

With the above, the streamlined interface can then be used as normal.
E.g.\ by calling the \ttt{hoppetEvolve(...)} function to fill the PDF
table and \ttt{hoppetEval(x,Q,f)} to evaluate the PDF at a given $x$
and $Q$.
Note that the \ttt{f} array in the latter call must be
suitably large, e.g.\ \ttt{f(-6:11)} for a PDF with leptons (numbering
is shifted by $+6$ in \CPP).
Examples of the streamlined interface being used with QED evolution
can be found in
\begin{quote}
  \masterlink{examples/f90/tabulation_example_qed_streamlined.f90}\\
  \masterlink{examples/cpp/tabulation_example_qed.cc}
\end{quote}

\subsection{Implementation of the QED extension}
\label{sec:qed-implementation}

\myparagraph{QED coupling}

A first ingredient is the setup of the QED coupling object, defined in
\ttt{module qed\_coupling}:
\begin{lstlisting}
  type qed_coupling 
     real(dp) :: m_light_quarks
     real(dp) :: mc, mb, mt
     integer  :: n_thresholds
     integer  :: nflav(3, 0:n_thresholds) ! first index: 1 = nleptons, 2=ndown, 3=nup
     real(dp) :: thresholds(0:n_thresholds+1)
     real(dp) :: b0_values(n_thresholds)
     real(dp) :: alpha_values(n_thresholds)
  end type qed_coupling
\end{lstlisting}
This is initialized through a call to 
\begin{lstlisting}
  subroutine InitQEDCoupling(coupling, m_light_quarks, &
   &                           m_heavy_quarks(4:6) [,value_at_scale_0])
    type(qed_coupling), intent(out) :: coupling
    real(dp),           intent(in)  :: m_light_quarks, m_heavy_quarks(4:6)
    real(dp), optional, intent(in)  :: value_at_scale_0 ! defaults to alpha_qed_scale_0
    [...]
  end subroutine InitQEDCoupling
\end{lstlisting}
It initialises the parameters relevant to the QED coupling and its
running.  The electromagnetic coupling at scale zero is set by default
to its PDG Thomson value~\cite{ParticleDataGroup:2022pth} value, unless the optional
argument \ttt{value\_at\_scale\_0} is provided, in which case the
latter is taken.

The running is performed at leading order level, using seven
thresholds: a common effective mass for the three light quarks
(\ttt{m\_light\_quarks}), the three lepton masses (hard-coded to their
2025 PDG values~\cite{ParticleDataGroup:2024cfk} in the
\ttt{src/qed\_coupling.f90} file), and the three masses of the heavy
quarks (\ttt{m\_heavy\_quarks(4:6)}).
The common value of the light quark masses is used to mimic the
physical evolution in the region $0.1\GeV \lesssim \mu \lesssim
1\GeV$, which involves hadronic states.
Using a value of $0.1055\GeV$ generates QED coupling values at the
masses of the $\tau$ lepton ($1/133.444$) and $Z$-boson ($1/127.938$)
that agree to within relative $\sim 1\times 10^{-4}$ (and $1\sigma$) accuracy with the
$\MSbar$ values ($1/(133.450 \pm 0.008)$ and $1/(127.930 \pm 0.006)$)
from the ``Electroweak and constraints on New Physics'' section of the
2024 Particle Data Group review~\cite{ParticleDataGroup:2024cfk}.

%
%

The quark and lepton masses are used to set all thresholds where the
fermion content changes.
The values of the thresholds are contained in an array
\ttt{threshold(0:8)}.
The \ttt{threshold(1:7)} entries are active thresholds, while
\ttt{threshold(0)} is set to zero and \ttt{threshold(8)} to an
arbitrary large number (currently $10^{200}$).
For a given $Q$, the code identifies the index \ttt{i} such that
\ttt{threshold(i-1)}$<Q<$ \ttt{threshold(i)}.
The flavour content at a given $Q$ is then accessible through the
integer array \ttt{nflav(3,0:n\_thresholds)}, where the
\ttt{nflav(1:3,i)} entries indicate respectively the number of
leptons, down-type, and up-type quarks at a given $Q$.
The \ttt{nflav(:,:)} array is then used to compute the
$\beta_{0,\rm QED}$ function \ttt{b0\_values(1:n\_thresholds)} at the
seven threshold values and this is finally used to compute the value
of the QED coupling \ttt{alpha\_values(1:n\_thresholds)} at the
threshold values.
The function \ttt{Delete(qed\_coupling)} is also provided for
consistency with general \hoppet conventions, although in this case it does nothing.
After this initialisation, the function \ttt{Value(qed\_coupling,mu)} returns the QED coupling at
scale $\mu$.

\myparagraph{QED splitting matrices}

The QED splitting matrices are stored in the object

\begin{lstlisting}
  type qed_split_mat
     type(qed_split_mat_lo)   :: lo
     type(qed_split_mat_nlo)  :: nlo
     type(qed_split_mat_nnlo) :: nnlo
  end type qed_split_mat
\end{lstlisting}
defined in \ttt{qed\_objects.f90}. 
This contains the LO, NLO and NNLO splitting matrices 
\begin{lstlisting}
  ! a leading-order splitting matrix (multiplies alpha/2pi)
  type qed_split_mat_lo
     type(grid_conv) :: Pqq_01, Pqy_01, Pyq_01, Pyy_01
     integer         :: nu, nd, nl, nf
  end type qed_split_mat_lo

  ! a NLO splitting matrix (multiplies (alpha alpha_s)/(2pi)^2)
  type qed_split_mat_nlo
     type(grid_conv) :: Pqy_11, Pyy_11, Pgy_11
     type(grid_conv) :: Pqg_11, Pyg_11, Pgg_11
     type(grid_conv) :: PqqV_11, PqqbarV_11, Pgq_11, Pyq_11
     integer         :: nu, nd, nl, nf
  end type qed_split_mat_nlo

  ! a NNLO splitting matrix (multiplies (alpha/(2pi) )^2) 
  ! contains only Plq splitting!
  type qed_split_mat_nnlo 
     type(grid_conv) :: Plq_02
     integer         :: nu, nd, nl, nf
  end type qed_split_mat_nnlo
\end{lstlisting}     
%
Above, \ttt{y} denotes a photon and the pairs of integers \ttt{01},
\ttt{11} and \ttt{02} denote the orders in the QCD and QED couplings,
respectively. Besides the number of quarks \ttt{nf}, these splitting
matrices also need the number of up-type (\ttt{nu}) and down-type
quarks (\ttt{nd}) separately, and the number of leptons (\ttt{nl}).
Note that the splitting functions of order $\alpha$ (i.e.\ \ttt{01})
for the leptons are simply obtained from the ones
involving quarks by adjusting colour factors and couplings.

A call to the subroutine 
\begin{lstlisting}     
  subroutine InitQEDSplitMat(grid, qed_split)
    use qed_splitting_functions
    type(grid_def),      intent(in)    :: grid
    type(qed_split_mat), intent(inout) :: qed_split
    [...]
  end subroutine InitQEDSplitMat
\end{lstlisting}     
initializes the \ttt{qed\_split\_mat} object \ttt{qed\_split} and sets all QED
splitting functions on the given \ttt{grid}.
The above QED objects can be used for any sensible value of the
numbers of flavours, on the condition that one first registers
the current number of flavours with a call to
\begin{lstlisting}
  QEDSplitMatSetNf(qed_split, nl, nd, nu)
\end{lstlisting}
where \ttt{nl}, \ttt{nd} and \ttt{nu} are respectively the current
numbers of light leptons, down-type and up-type quarks.
In practice, this is always handled internally by the QED-QCD
evolution routines, based on the thresholds encoded in the QED
coupling.\footnote{While the QCD splitting functions are initialised
  and stored separately for each relevant value of $n_f$, in the QED
  case the parts that depend on the numbers of flavours are separated
  out.
  Only when the convolutions with PDFs are performed are the relevant
  $n_f$ and electric charge factors included.}
The one situation where a user would need to call this routine
directly is if they wish to manually carry out convolutions of the
QED splitting functions with a PDF.

Subroutines \ttt{Copy} and \ttt{Delete} are also provided for the
\ttt{qed\_split\_mat} type. As in the pure QCD case, convolutions with
QED splitting functions can be represented by the \ttt{.conv.} operator or
using the product sign \ttt{*}.

\myparagraph{PDF arrays with photons and leptons}

A call to the subroutine
\begin{lstlisting}  
  subroutine AllocPDFWithPhoton(grid, pdf)
    type(grid_def), intent(in) :: grid
    real(dp),       pointer    :: pdf(:,:)
    [...]
  end subroutine AllocPDFWithPhoton
\end{lstlisting}
allocates PDFs (\ttt{pdf}) including photons, while a call to
\begin{lstlisting}  
  subroutine AllocPDFWithLeptons(grid, pdf)
    type(grid_def), intent(in) :: grid
    real(dp),       pointer    :: pdf(:,:)
    [...]
    end subroutine AllocPDFWithLeptons
\end{lstlisting}
allocates PDFs including both photons and leptons.
The two dimensions of the \ttt{pdf} refer respectively to
the index of the $x$ value in the \ttt{grid}, and to the flavour index.
The flavour indices for photons and leptons 
are
given by
\begin{lstlisting}  
  integer, parameter, public :: iflv_photon   =  8
  integer, parameter, public :: iflv_electron =  9
  integer, parameter, public :: iflv_muon     = 10
  integer, parameter, public :: iflv_tau      = 11
\end{lstlisting}
where each \ttt{pdf(:,9:11)} contains the sum of a lepton and
anti-lepton flavour (which are identical). Note that if one were to
extend the calculation of lepton PDFs to higher order in $\alpha$,
then an asymmetry in the lepton and anti-lepton distribution would
arise, due to the \ttt{Plq\_03} splitting function.
%
In fact, at that order, there are also graphs with three
electromagnetic vertices on the quark line and three on the lepton line, that change sign if the lepton line is charge-conjugated.\footnote{Analogous terms appear in the non-singlet three-loop splitting functions~\cite{NNLO-NS}.}
In that case it would become useful
to have separate indices for leptons and anti-leptons.


The subroutine \ttt{AllocPDFWithPhotons} allocates the \ttt{pdf}
array with the flavour index from -6 to 8,
while in the subroutine \ttt{AllocPDFWithLeptons},
the flavour index extends from -6 to 11.

\myparagraph{PDF tables with photons and leptons}

Next one needs to prepare a \ttt{pdf\_table} object forming the
interpolating grid for the evolved PDF's. We recall that the
\ttt{pdf\_table} object contains an underlying array \ttt{pdf\_table\%tab(:,:,:)},
where the first index loops over $x$ values, the second loops over
flavours and the last loops over $Q^2$ values.

This is initialized by a call to 
\begin{lstlisting}
  subroutine AllocPdfTableWithLeptons(grid, pdftable, Qmin, Qmax & 
  & [, dlnlnQ] [, lnlnQ_order] [, freeze_at_Qmin])
   use qed_objects
   type(grid_def),    intent(in)    :: grid
   type(pdf_table),   intent(inout) :: pdftable
   real(dp), intent(in)             :: Qmin, Qmax
   real(dp), intent(in), optional   :: dlnlnQ
   integer,  intent(in), optional   :: lnlnQ_order
   logical,  intent(in), optional   :: freeze_at_Qmin
   [...]
   end subroutine AllocPdfTableWithLeptons
\end{lstlisting}
that is identical to the one without photon or leptons, the only difference is that the
maximum pdf flavour index in \ttt{pdf\_table\%tab} now includes the photon and leptons.
An analogous subroutine \ttt{AllocPdfTableWithPhoton} includes the
photon but no leptons.

\myparagraph{Evolution with photons and leptons}

To fill a table via an evolution from an initial scale, one calls the subroutine 
\begin{lstlisting}
    subroutine EvolvePdfTableQED(table, Q0, pdf0, dh, qed_split, &
    &  coupling, coupling_qed, nloop_qcd, nloopqcd_qed, with_Plq_nnloqed)
       type(pdf_table),        intent(inout) :: table
       real(dp),               intent(in)    :: Q0
       real(dp),               intent(in)    :: pdf0(:,:)
       type(dglap_holder),     intent(in)    :: dh
       type(qed_split_mat),    intent(in)    :: qed_split
       type(running_coupling), intent(in)    :: coupling
       type(qed_coupling),     intent(in)    :: coupling_qed
       integer,                intent(in)    :: nloop_qcd, nloopqcd_qed
       logical,  optional,     intent(in)    :: with_Plq_nnloqed
       [...]
    end subroutine EvolvePdfTableQED   
\end{lstlisting}
where \ttt{table} is the output, \ttt{Q0} the initial scale and
\ttt{pdf0} is the PDF at the initial scale.
We recall that the lower and upper limits on 
scales in the table are as set at initialisation time for the table.
When \ttt{nloopqcd\_qed} is set to 1 (0) mixed QCD-QED effects are
(are not) included in the evolution.
Setting the variable \ttt{with\_Plq\_nnloqed=.true.}\ includes also the NNLO $P_{lq}$
splittings in the evolution.

To perform the evolution \ttt{EvolvePdfTableQED} calls the routine 
\begin{lstlisting}
    subroutine QEDQCDEvolvePDF(dh, qed_sm, pdf, coupling_qcd, coupling_qed,&
    &                     Q_init, Q_end, nloop_qcd, nqcdloop_qed, with_Plq_nnloqed)
       type(dglap_holder),     intent(in), target :: dh
       type(qed_split_mat),    intent(in), target :: qed_sm
       type(running_coupling), intent(in), target :: coupling_qcd
       type(qed_coupling),     intent(in), target :: coupling_qed
       real(dp),               intent(inout)      :: pdf(0:,ncompmin:)
       real(dp),               intent(in)         :: Q_init, Q_end
       integer,                intent(in)         :: nloop_qcd
       integer,                intent(in)         :: nqcdloop_qed
       logical,  optional,     intent(in)         :: with_Plq_nnloqed
       [...]
    end subroutine QEDQCDEvolvePDF
\end{lstlisting}
Given the \ttt{pdf} at an initial scale \ttt{Q\_init}, it
evolves it to scale \ttt{Q\_end}, overwriting the \ttt{pdf} array. 
In order to get interpolated PDF values from the table we use the
\ttt{EvalPdfTable\_*} calls, described in
\ifreleasenote
Section~7.2 of Ref.~\cite{Salam:2008qg}.
\fi
However the \texttt{pdf} array that is passed as an argument and that
is set by those subroutines should range not from \texttt{(-6:6)} but
instead from \texttt{(-6:8)} if the PDF just has photons and
\texttt{(-6:11)} if the PDF also includes leptons.\footnote{Index $7$
  is reserved in \hoppet to hold information about the flavour
  representation of PDFs.
  Ideally PDFs with the full set of flavours would be represented by an
  underlying type with a 2D array for storing the $x$-dependence of
  the different PDFs, supplemented with an index for the representation.
  However when we explored this option during v1 development we found
  that this compromised our ability to maintain speed while writing
  operations of PDFs such as \ttt{a=b+d*c}.
  We thus opted to work with plain 2D arrays, encoding the flavour
  structure with a specific signature for the $x$-dependence of index
  7.
  This is not ideal, however changing it would come at the price of
  breaking backwards compatibility, and one would still need to
  explore whether, with today's Fortran features, one could find a
  good solution without a speed impact.}

Note that at the moment, when QED effects are included, cached
evolution is not supported.

\section{Python interface}
\label{sec:pyinterface}
From version~2.0.0, \hoppet also includes a Python interface to the
most common evolution routines. It currently includes exactly the same
functionality as the streamlined interface, and can be called in much
the same way. The names of functions and routines are the same as in
the streamlined interface, but stripped of the \ttt{hoppet}
prefix. The interface can be obtained from \ttt{PyPi} by invoking
\begin{lstlisting}
  pip install hoppet
\end{lstlisting}
Alternatively the interface can be built by CMake with
\ttt{-DHOPPET\_BUILD\_PYINTERFACE=ON} (cf.\ Section~\Ref{sec:cmake}
for details on building with CMake). In both cases the interface
can be imported into a \ttt{Python} instance through \ttt{import
  hoppet}.
For a simple tabulation example, the user should take a look
at
\repolink{examples/python/tabulation_example.py}{examples/python/tabulation\_example.py}
(and in the same directory for a number of other illustrative examples
of how to use the interface, including in conjunction with
LHAPDF~\cite{LHAPDF}). The interface uses SWIG (Simplified Wrapper and
Interface Generator) which therefore needs to be available on the
system if building with CMake. It can be installed through most
package managers (e.g.\ \texttt{apt install swig}, \texttt{brew
  install swig}) but can also be obtained from the SWIG GitHub
\href{https://github.com/swig}{repository}.

One significant difference between the Python interface and the
streamlined interface is that Python does not provide native support
for pointers.
A number of \CPP routines
fill an array of PDF flavours that is passed as a pointer
argument.
Instead in Python, those routines return a
Python list directly.
For instance, where in \CPP{}
one might have the call
\begin{lstlisting}
  #include "hoppet.h"
  [...]
  double x = ...;
  double Q = ...;
  double pdf[13];
  [...]
  hoppetEval(x, Q, pdf);
\end{lstlisting}
instead in Python one would have
\begin{lstlisting}
  import hoppet as hp
  [...]
  x = ...
  Q = ...
  pdf = hp.Eval(x,Q)
\end{lstlisting}
This is relevant not just for evaluation of PDFs, but also in setting
initial conditions. 
For example the \ttt{Assign}, \ttt{CachedEvolve}, and
\ttt{Evolve} routines should be passed a function
of 
\ttt{x} and \ttt{Q} that returns an object that corresponds to array
of flavours (it can be a \ttt{numpy}~\cite{Harris:2020xlr} array of a
Python list; see the \ttt{hera\_lhc(x,Q)} function in
\repolink{examples/python/tabulation_example.py}{tabulation\_example.py}
for an explicit example).

In addition to the examples provided in
\repolink{examples/python/}{examples/python/} we have also developed a small
tool that loads a grid from LHAPDF at an initial scale and evolves it
with \hoppet{} over a large range of $Q$. The resulting grids are then
compared between \hoppet{} and LHAPDF to determine the relative
accuracy of the LHAPDF grids. The tool, along with some documentation,
can be found at
\url{https://github.com/hoppet-code/hoppet-lhapdf-grid-checker}.

\section{CMake build system}
\label{sec:cmake}

In v1.x, \hoppet used a hand-crafted \ttt{./configure} script followed
by \ttt{make [install]}.
As of v2
\hoppet uses CMake.\footnote{We thank Andrii Verbitskyi
  for providing much of this system.
  Support for the old build system was retained in the briefly lived
  2.0.x series, but was retired in version 2.1.0 owing to the need to
  support compilation with both Fortran and \CPP, the latter for the
  \ttt{libome} library.}

For a typical user it will be enough to invoke the following lines
from the main directory 
\begin{lstlisting}
  cmake -S . -B build
  cmake --build build [-j<num_cores>]
  ctest --test-dir build [-j<num_cores>]
  cmake --install build
\end{lstlisting}
%
%
This will compile and install \hoppet, along with the streamlined interface.
Note that \ttt{cmake --install build} will typically
install in a location that requires root privileges, unless a user has
specified a custom prefix (through
\ttt{-DCMAKE\_INSTALL\_PREFIX=/install/path}). A number of options can
be passed to CMake. They are documented in
\repolink{CMakeLists.txt}{CMakeLists.txt} and can be printed on screen
by a call to \ttt{cmake -LH ..} from the \ttt{build} directory.

Of particular note to most users are 
\begin{lstlisting}
  option(HOPPET_USE_EXACT_COEF    "Use exact coefficient functions"  OFF)
  option(HOPPET_BUILD_EXAMPLES    "Build examples" ON)
  option(HOPPET_ENABLE_TESTING    "Enable testing. Requires building the examples." ON)
  option(HOPPET_BUILD_BENCHMARK   "Build benchmark." ON)
  option(HOPPET_BUILD_PYINTERFACE "Build python interface." OFF)
  option(HOPPET_ENABLE_FPES       "Enable trapping for the floating point exceptions." OFF)
  option(HOPPET_BUILD_TESTDIR     "Enable build of the code in the test directory." OFF)
\end{lstlisting}
They can be set in the usual \ttt{cmake} way, e.g.\
\begin{lstlisting}
  cmake -S . -B build -DHOPPET_USE_EXACT_COEF=ON
\end{lstlisting}
to compile \hoppet with the exact coefficient functions.
Note that the Python interface is not compiled by default, because we
anticipate that users of Python will prefer to obtain \hoppet through
\ttt{pip install hoppet}.

Users should be aware that this release of \hoppet makes use of
features introduced in Fortran~2008, for example its \ttt{abstract
  interface}, and hence a Fortran~2008 compliant compiler is now
needed. The code has been tested to compile and run with
\texttt{gfortran v10.5.0} and later, and the 2025 version of the Intel
compiler \texttt{ifx}.
If multiple Fortran compilers are available, a specific one can be
chosen with the \ttt{-DCMAKE\_Fortran\_COMPILER=...} option.

\section{Saving LHAPDF grids}
\label{sec:lhapdf}
A minor new feature of this release is the possibility to save a
\hoppet{} table in the form of an LHAPDF6~\cite{LHAPDF} grid.
The main routine has the following structure
\begin{lstlisting}
  subroutine WriteLHAPDFFromPdfTable(table, coupling, basename, pdf_index, &
                                 & iy_increment, flav_indices, flav_pdg_ids, flav_rescale)                                
    use qed_objects
    type(pdf_table),        intent(in) :: table
    type(running_coupling), intent(in) :: coupling
    character(len=*),       intent(in) :: basename
    integer,                intent(in) :: pdf_index
    integer,  optional,     intent(in) :: iy_increment
    integer,  optional,     intent(in) :: flav_indices(:), flav_pdg_ids(:)
    real(dp), optional,     intent(in) :: flav_rescale(:)
\end{lstlisting}
A user has to provide a \ttt{table} and associated \ttt{coupling}
object along with a string \ttt{basename} and the \ttt{pdf\_index} as
needed by LHAPDF.
If \ttt{pdf\_index} is equal to 0 then the routine
outputs the contents of the table in \ttt{basename\_0000.dat} and
writes a template \ttt{basename.info} again in LHAPDF
format.
\hoppet{} fills most of the entries in the \ttt{.info}
file, but a few need to be edited manually by the user.
For any other value of \ttt{pdf\_index}, only the corresponding
\ttt{.dat} file gets written.

By default, the code uses the same grid spacing as in the internal
\hoppet{} table (\ttt{iy\_increment = 1}) and prints all the possible
flavours of the PDF, even if they are zero.
The user can overwrite
this default behaviour by providing an array of flavours and their pdg
values.
If \ttt{iy\_increment > 1} a coarser grid is provided by
skipping over \ttt{iy\_increment - 1} points in the grid. The
\ttt{flav\_rescale} is currently needed for the lepton PDF which is
provided as the sum over flavour and anti-flavour and therefore needs
an extra factor half.
The \ttt{flav\_rescale} array only needs to be
provided if the user is also providing the array of flavours and pdf
values.

Finally the routine can also be accessed from the streamlined
interface for \CPP{} or Python usage. In this case the routine is
significantly simplified and only takes \ttt{basename} and the
\ttt{pdf\_index} arguments, for instance in \CPP{} like this
\begin{lstlisting}
  #include "hoppet.h"
  #include <string>
  ...
  int main () {
    ...
    const std::string basename = ...;
    const int pdf_index = ...;
    hoppetWriteLHAPDFGrid(basename, pdf_index);
    ...
  }
\end{lstlisting}
The routine writes the contents of the streamlined interface
\ttt{tables(0)} and hence requires that this object has been filled
either through a call to \ttt{hoppetAssign} or \ttt{hoppetEvolve}.

\section{Updated performance studies}
\label{sec:v2-performance}

In this section we present some updated performance studies relative
to
\ifreleasenote
Section~9 of Ref.~\cite{Salam:2008qg},
\fi
mainly reflecting updated hardware and compilers of 2025, but also the
standard nested grid choice that is obtained with the streamlined
interface or, from the modern Fortran interface by calling
\ttt{InitGridDefDefault(grid, dy, ymax[, order])}, with the default
choice of interpolation \ttt{order=-6}.
\ifreleasenote
The \ttt{InitGridDefDefault(...)} routine is new relative to v1 and sets up
the following grid
\begin{lstlisting}
  call InitGridDef(gdarray(4),dy/27.0_dp, 0.2_dp, order)
  call InitGridDef(gdarray(3),dy/9.0_dp,  0.5_dp, order)
  call InitGridDef(gdarray(2),dy/3.0_dp,  2.0_dp, order)
  call InitGridDef(gdarray(1),dy,         ymax  , order)
  call InitGridDef(grid,gdarray(1:4),locked=.true.)
\end{lstlisting}
Users will usually only need a different choice if they plan studies
at $x$ very close to $1$ or if they wish to explore fine optimisation of
grid choices.
\fi

We split our study here into two parts: the accuracy of PDF evolution
and tabulation (section~\ref{sec:acc-pdf-evol}) and PDF evaluation
(section~\ref{sec:fastpdf}).
The latter also outlines new functionality for choosing interpolation
orders differently in the PDF evaluation versus the PDF evolution and
it includes comparisons to LHAPDF.


\subsection{PDF evolution and tabulation}
\label{sec:acc-pdf-evol}

\begin{figure}[htbp]
    \centering
    \begin{subfigure}{0.49\textwidth}
        \centering
        \includegraphics[width=\textwidth,page=2]{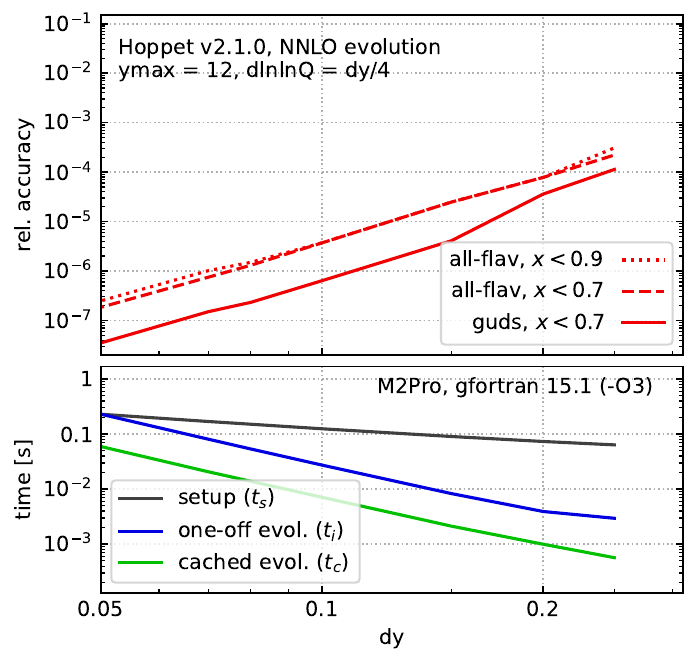}
        \caption{}
        \label{fig:main-sub1}
    \end{subfigure}
    \hfill
    \begin{subfigure}{0.49\textwidth}
        \centering
        \includegraphics[width=\textwidth,page=5]{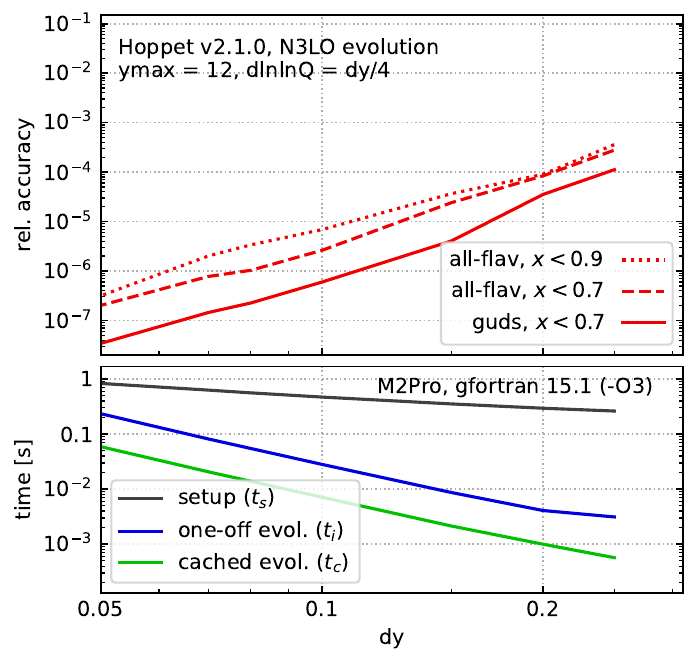}
        \caption{}
        \label{fig:main-sub2}
    \end{subfigure}
    \caption{Accuracy (top) and timing (bottom) versus \texttt{dy} at
      (a) NNLO and (b) N$^3$LO in \hoppet{} v2.1.0.
      The accuracy corresponds to the worst fractional accuracy for any flavour,
      at any $x$ value below the corresponding limit, as described in
      the text.
      The timings were
      obtained on an \texttt{M2Pro} with \texttt{gfortran v15.1} and
      \texttt{-O3} optimisation. We have further used \texttt{ymax = 12} and
      \texttt{dlnlnQ = dy/4}.
    }
    \label{fig:main}
\end{figure}

The studies are performed using the initial condition of
Ref.~\cite{Dittmar:2005ed} as detailed in
\ifreleasenote
Section~9.1 of Ref.~\cite{Salam:2008qg},
\fi
evolved in the VFNS scheme.
At NNLO we use the parametrised splitting
functions and mass thresholds, as per \hoppet defaults.
To
assess the accuracy we first create a reference run with a very high
density grid. We then run \hoppet{} with different values of grid
spacing \texttt{dy}, keeping \texttt{dlnlnQ = dy/4}.
The accuracy is then computed by looking at the largest relative
deviation from the reference run across either all flavours (all-flav)
or the light flavours (guds).
The full tabulation covers the range $10^{-5} < x < 1$,
$2 < Q < 10^4\, \text{GeV}$.\footnote{For tabulations extending to
  significantly smaller $x$ values, it can be advantageous to take a
  smaller \texttt{dlnlnQ} choice, e.g.\ \texttt{dlnlnQ = dy/8},
  because of the steep $\ln Q$ derivative of the parton
  distributions at the smallest $x$ values.}
At $x$ values close to $1$, the numerical precision degrades because the
parton distribution functions become a very steep function of $\ln
x$.
However, the parton distributions have small values there and so we
carry out our precision study with two potential upper limits on the
$x$ value being probed, $x < 0.9$ and $x < 0.7$.
We also exclude PDF flavours in the $x$ and $Q$ vicinity of any sign
change, as per Ref.~\cite{Salam:2008qg}.

The results for the accuracy study can be seen in the top panels of
Figures~\ref{fig:main-sub1}--\ref{fig:main-sub2} at NNLO and N$^3$LO
respectively, as a function of \texttt{dy}.
At NNLO, the accuracy comes out similar to previous
versions of \hoppet.
With \texttt{dy =
  0.2} one obtains a relative accuracy of $10^{-4}$ across all
flavours in the range $x < 0.9$.
At the finest grid spacing \texttt{dy = 0.05}, a relative accuracy of
few times $10^{-7}$ can be achieved, good enough for precise benchmark
comparisons as were for instance carried out in
Refs.~\cite{Dittmar:2005ed,Bertone:2024dpm}.
For comparison,
the recent benchmark of aN$^3$LO codes in
Ref.~\cite{Cooper-Sarkar:2024crx} reaches a relative precision of a
few times $10^{-4}$ at best (cf.\ the gluon PDF at $x = 10^{-2}$ in
Table~\ref{tab:n3lo-evolve-nf4} and Table 1 of
Ref.~\cite{Cooper-Sarkar:2024crx} which differ by $\sim 8\cdot 10^{-4}$.)

The scaling of the precision is roughly consistent with a power law in
\texttt{dy}.
In particular the Runge-Kutta algorithm for the $Q^2$ evolution is
expected to yield an error proportional to $\texttt{dlnlnQ}^4$, which,
given our choice of $\ttt{dlnlnQ} = \ttt{dy}/4$
translates into a behaviour $\sim \texttt{dy}^4$ in
Fig.~\ref{fig:main}.
The observed scaling is, if anything, slightly better than this, given
the factor of $1000$ improvement in accuracy when reducing \texttt{dy}
by a factor of $5$ from $0.2$ to $0.04$.
The precise scaling depends on whether it is Runge-Kutta or the
splitting function grid representation that dominates the error.
At N$^3$LO we observe a similar level of accuracy as at NNLO, except
for a slight worsening associated with the heavy-flavour components.



For the timing studies we again run \hoppet{} for different values of
\texttt{dy}, on an \texttt{M2Pro} (MacOS 15.6.1) with \texttt{gfortran v15.1} and
\texttt{-O3} optimisation.
%
As discussed in \cite{Salam:2008qg}, the time spent in \hoppet for a
given analysis can be expressed as follows, depending on whether or
not one carries out cached evolution (pre-evolution):
\begin{subequations}
  \label{eq:timing-v2}
  \begin{align}
    t_\text{no pre-ev}   &= t_s + n_\alpha t_\alpha + n_i (t_i  + n_{xQ}\, t_{xQ})\,,\\
    t_\text{with pre-ev} &= t_s + n_\alpha (t_\alpha + t_p) + n_i (t_c + n_{xQ}\,
    t_{xQ})\,,
  \end{align}
\end{subequations}
where $t_s$ is the time for setting up the splitting functions and
mass threshold functions, $n_\alpha$ is the number of different
running couplings that one has, $t_\alpha$ is the time for
initialising the coupling,
$n_i$ is the number of PDF initial conditions that one wishes to
consider, $t_i$ is the time to carry out the tabulation and evolution
for a single initial condition, $n_{xQ}$ is the number of points in
$x,Q$ at which one evaluates the full set of flavours once per PDF
initial condition; in the case with cached evolution, $t_p$ is the
time for preparing a cached evolution and $t_c$ is the time for
performing the cached evolution. Finally, $t_{xQ}$ is the time it takes
to evaluate the PDFs at a given value of $(x,Q)$ once the tabulation
has been performed.

Here we focus on $t_s$, $t_i$, and $t_c$. The results can be seen in the
bottom panels of Figures~\ref{fig:main-sub1}--\ref{fig:main-sub2} at NNLO and
N$^3$LO respectively.
The expected scaling is
$t_i, t_c \sim (\text{\texttt{dy}}^2 \text{\texttt{dlnlnQ}})^{-1}$, which for our
choice of $\texttt{dlnlnQ} \propto\texttt{dy}$ reduces to
$1/\text{\texttt{dy}}^3$.
That is consistent with what is seen in the plot.
Turning to the setup time, at NNLO 
$t_s \sim 60-300\,\mathrm{ms}$ dominates over the evolution time across
almost all \texttt{dy} values that we study.
It scales slightly more slowly than $t_s \sim 1/\texttt{dy}$.

We note that when using cached evolution, the evolution time $t_c$ reaches as
little as $1\,\mathrm{ms}$ for \texttt{dy = 0.2}.
Comparing these numbers to those of  
\ifreleasenote
Table~2 of Ref.~\cite{Salam:2008qg},
\fi
which were obtained with 2008 hardware and compilers, we see a
speed-up of roughly a factor 10, which we attribute mainly to
improvements in the hardware.

At \ntlo  evolution times ($t_i$ and $t_c$) are essentially identical to the
NNLO case: for $t_i$ the only extra operation that is needed is the
addition of the \ntlo splitting function to the lower-order ones and
at each $Q$ value, this involves $\order{N}$ operations for a $y$-grid
of size $N$, while the convolution itself involves $\order{N^2}$
operations.
For the cached evolution, there is no additional penalty, because the
\ntlo contributions are already included in the cached evolution
operators.
The initialisation times are somewhat larger, in a range from 250\,ms
to 1\,s.
The longer time is associated both with the approximate \ntlo
splitting functions and the mass threshold functions of
Ref.~\cite{BlumleinCode}, using the \ttt{n3lo\_nfthreshold\_libOME} option.
In practice the initialisation time remains adequate for most
interactive work.
In any long-running application where \hoppet either has to evolve or
access the evolved tables many times, the initialisation time is
insignificant.
%



\subsection{Fast PDF access}
\label{sec:fastpdf}

Here we detail updates for faster PDF access within the
modern Fortran and streamlined interfaces (including the Python interface).
%
%
In earlier versions of \hoppet, the interpolation was carried out by a
single routine that could flexibly handle any choice of $y$ and $Q$
interpolation orders up to some hard-coded maximum.
As of v2.0.0, a number of interpolation-order choices now have
dedicated code, which makes it easier for the compiler to optimise the
underlying assembly, e.g.\ with loop unrolling, giving speed gains of
almost a factor of three.
Additionally new functionality allows the user to modify
the interpolation order of the \hoppet{} grids, trading accuracy
versus speed.


Specificaly, the user can now globally override any table-specific interpolation order
settings by calling one of
\begin{lstlisting}
  call hoppetSetYLnlnQInterpOrders(yorder, lnlnQorder)  ! streamlined interface
\end{lstlisting}
or
\begin{lstlisting}
  call PdfTableOverrideInterpOrders(yorder, lnlnQorder) ! F90 interface
\end{lstlisting}
A value of $\ttt{order}=2$ corresponds to quadratic interpolation, $3$ to
cubic interpolation, etc.\footnote{The \ttt{(yorder, lnlnQorder)}
  choices with dedicated code are $(2,2)$, $(3,3)$, $(4,4)$, $(5,4)$
  and $(6,4)$.}
The default interpolation order is quartic in the \ttt{lnlnQ}
direction, and \ttt{|grid\%order|-1} in the \ttt{y} direction (bounded
to be between 3 and 9 if outside that range).
These rather high interpolation orders help ensure good accuracy in a normal
\hoppet{} run even with $\ttt{dy=0.2}$, but come with a speed penalty
because of the larger number of operations.
However, if PDF evaluation represents a significant fraction of the
time for a user's code, the user can choose to lower the interpolation
order and still retain good accuracy by reducing \ttt{dy} and
\ttt{dlnlnQ}.

\begin{figure}[tbp]
    \centering
        \includegraphics[width=0.49\textwidth,page=3]{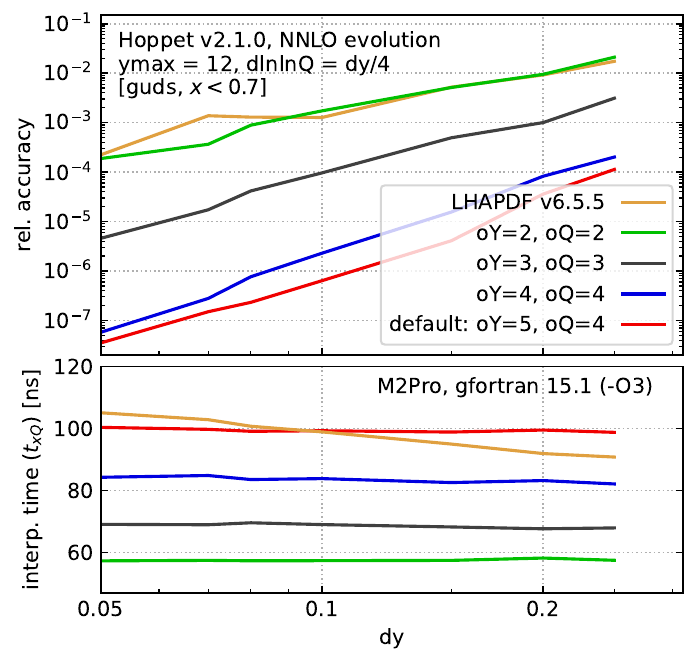}
    \caption{
      Study of the impact of the choice of the $y$ and $Q$ interpolation
      orders (\ttt{oY}, \ttt{oQ}) on accuracy (upper panel) and speed
      (lower panel) for evaluating
      the PDF table at a given $y$ and $Q$ point.
      Note that the interpolation order that enters the splitting
      functions remains $6$ as in Fig.~\ref{fig:main}.
      The timing corresponds to the \ttt{EvalPdfTable\_yQ(table,y,Q,vals)} modern
      Fortran call.
      The timings have been
      obtained on an \texttt{M2Pro} with \texttt{gfortran v15.1} and
      \texttt{-O3} optimisation.
      Calling \ttt{EvalPdfTable\_xQ(table,x,Q,vals)}, or the corresponding
      functions in the streamlined interfaces, adds about $5\ns$.
      We have further used \texttt{ymax = 12} and
      \texttt{dlnlnQ = dy/4}.
      Note that our accuracy definition is to show the \emph{worst}
      accuracy across any $[guds]$ flavour and $x$ value in range.
      In practice, most points have a significantly higher
      accuracy. We also show results obtained by interpolating our
      grids with LHAPDF 6.5.5, calling it from its native \CPP
      interface to maximise its speed.  }
        \label{fig:interp-accuracy}
\end{figure}

Fig.~\ref{fig:interp-accuracy} shows the accuracy and timing as a
function of \ttt{dy} for different interpolation order choices.
It compares evolution as in Fig.~\ref{fig:main}, followed by
interpolation with a given order.
It illustrates the significant loss in accuracy when decreasing
the interpolation order below $4$.
For orders of $4$ or higher, the limitation on accuracy is, however,
no longer just the interpolation order during PDF evaluation, but also
the orders that appear in the grid representation of the splitting
functions and of the $Q$ evolution when producing the original
table.\footnote{Recall that the former can be controlled with the
  \ttt{order} argument when setting up the grid.
  Currently, the latter is hard-coded as part of the Runge-Kutta
  evolution routines (cf.\ above).
}
The time that is shown in the lower panel corresponds to the
evaluation of all flavours in one go, via the
\ttt{EvalPdfTable\_yQ(...)} routine.
It ranges from $60\ns$ to $100\ns$ in going from the $(2,2)$ order
combination to $(5,4)$.
It is largely independent of \ttt{dy}.
Calls to evaluate a single flavour are only $2.5{-}3$ times faster,
because of overheads associated with identifying the grid location and
calculation interpolation coefficients, which are independent of the
number of flavours one evaluates.

Fig.~\ref{fig:interp-accuracy} also shows results with LHAPDF, calling
it from its native \CPP to maximise its speed.
We generate an LHAPDF grid for our standard benchmark initial
conditions, with a given \ttt{dy} spacing, cf.\
Section~\ref{sec:lhapdf}, and then examine the difference between the
LHAPDF evaluation and our high-accuracy reference.
The upper panel of Fig.~\ref{fig:interp-accuracy} shows the accuracy
with the same definition as used in our other performance results,
i.e.\ the worst relative accuracy observed anywhere for $x < 0.7$,
across any of the g,u,d,s flavours, as in the solid lines of
Fig.~\ref{fig:main}.
For most flavours and most of the $x$ region, the accuracy is better
than shown.
Interestingly, the LHAPDF accuracy is comparable to \hoppet's
quadratic interpolation.
At first sight this might be surprising given that LHAPDF uses cubic
interpolation: however it is our understanding that in the $x$
direction it uses a cubic \emph{spline}.
The spline effectively sacrifices one of the orders of accuracy in
function evaluation and instead uses it to ensure exact continuity of
the first derivative.
In contrast, \hoppet does not enforce continuity of the derivatives
but relies on the fact that for high interpolation order any
discontinuity of the derivatives will scale as a high power of the
grid spacing. 
Which choice is better may depend on the application. 
Concerning speed, we find that LHAPDF is somewhere in between our
$(4,4)$ and $(5,4)$ choices.

For use-cases where PDF evaluation speed is critical, one option is
to read in a PDF grid with LHAPDF, evaluate it at all \hoppet grid points and
then use \hoppet for the interpolation, using a fine grid spacing and
a $(2,2)$ interpolation choice.
We provide examples for how to do this in Fortran, \CPP, and Python:
\begin{itemize}
 \item \masterlink{examples/f90/with-lhapdf/lhapdf_to_hoppet.f90}: it
   uses the module \ttt{hoppet\_lhapdf} and the associated
   \ttt{subroutine LoadLHAPDF(name, imem)}. The module is in the same
   directory and can be copied over to a user's application;
 \item
   \masterlink{examples/f90/with-lhapdf/lhapdf_to_hoppet_allmembers.f90},
   same as above but the example shows how to read in and efficiently
   evaluate all
   LHAPDF members, with an illustration for computing the PDF uncertainty;
 \item \masterlink{examples/cpp/with-lhapdf/lhapdf_to_hoppet.cc}, which
   includes a routine called \ttt{void load\_lhapdf\_assign\_hoppet(const
     string \& pdfname, int imem=0)} which can be included directly in a
   user's application;
 \item \masterlink{examples/python/lhapdf_to_hoppet.py}, makes use
   of \ttt{hoppet.lhapdf.load()}, accessible using ``\ttt{from hoppet import lhapdf}''.
\end{itemize}
%
%
%
The PDF can then be evaluated in the usual
way through a call to \ttt{hoppetEval} and likewise for the coupling
through \ttt{hoppetAlphaS}.\footnote{%
  For the PDF, \hoppet just
  interpolates its table, which is filled directly from LHAPDF.
  For $\as$, \hoppet carries out its own evolution, which may have
  small differences from the LHAPDF $\as$, notably if the LHAPDF grid
  provides a coupling that is not an exact solution of the renormalisation
  group equation.
  Another subtlety concerns the top threshold.
  The LHAPDF grid files usually quote a physical top mass, and in the
  examples above, \hoppet reads the top mass and provides $\as$
  evolution at higher scales with $6$ light flavours.
  However in some LHAPDF sets, the top-mass is only intended to
  indicate the value of the top mass used in the PDF fit, not an $n_f = 5
  \to 6$ threshold in the evolution.
  The examples above all use a call similar to
  \ttt{if(!pdf->hasFlavor(6)) mt = 2.0*Qmax;} which manually overrides
  the top mass to a value outside the grid ranges if there is no top
  quark present in the set.
  %
}
All three examples also print the timings to the screen so that users
can check speed on their hardware. Note that these examples are not
fully general, and caution should be exercised when using them. Users
with more advanced requirements are invited to contact the \hoppet
authors for assistance.

\begin{table}[htbp]
  \centering
  \newcolumntype{C}{>{\centering\arraybackslash}p{3cm}}
  \begin{tabular}{|c|c|CCC|}
    \hline
    \ttt{yorder} & \ttt{lnlnQorder} & \multicolumn{3}{c|}{Time per \ttt{hoppetEval} or
                                        LHAPDF \ttt{xfxQ}/\ttt{EvolvePDF} call (ns)} \\
     \hline
      &                          & Fortran & \CPP   & Python \\
    5 & 4                        & 108     & 108    & 295 \\
    4 & 4                        &  92     &  93    & 278 \\
    3 & 3                        &  76     &  77    & 258 \\
    2 & 2                        &  63     &  63    & 244 \\ \hline\hline
    \multicolumn{2}{|l|}{\rule{0pt}{2.5ex}HOPPET 1.2.0}
                                 & 313     & 313    &  -- \\ 
    \multicolumn{2}{|l|}{\rule{0pt}{2.5ex}LHAPDF 6.5.5}
                                 &520 &  87    & 1645 \\
    \multicolumn{2}{|l|}{\rule{0pt}{2.5ex}LHAPDF 6.5.5 + patch}
                               &114 &  87    & 1028 \\
    \hline
  \end{tabular}
  \caption{Time for a call to
  \ttt{hoppetEval}, which evaluates all flavours at a given $x,Q$
  point.
  The rows show the timings with different
  interpolation orders.
  The results have been obtained with the \ttt{PDF4LHC21\_40} set, on an M2Pro with
  gfortran-15.1, Apple clang version 17.0.0, O3 optimisation and
  Python 3.12.7.
  The hoppet grid spacings are $\ttt{dy}=0.05$ and
  $\ttt{dlnlnQ}=\ttt{dy}/4$, with $\ttt{ymax} = 14$.
  %
  The table also shows the timings for the equivalent calls in
  LHAPDF~6.5.5.
  Version 6.5.5 of LHAPDF, in its Fortran (\ttt{EvolvePDF(...)},
  \ttt{lhapdf\_xfxq(...)}) and Python (\ttt{pdf.xfxQ(...)}) 
  interfaces, 
   loops over an underlying call to single-flavour evaluation
  (\ttt{double PDF::xfxQ(int id, double x, double q))}), 
  which is why it is significantly slower than the \CPP interface,
  which evaluates all flavours in one go.
  A proposed patch for the Fortran and Python interfaces has been submitted
  to LHAPDF with corresponding timings also indicated.
}
  \label{tab:interp_performance}
\end{table}

Table~\ref{tab:interp_performance} illustrates those timings for a range of
interpolation orders and across interfaces in different languages.
These tests have been carried out with the \ttt{PDF4LHC21\_40} set.
Again, they confirm that \hoppet with lower interpolation orders
can offer a speed gain relative to LHAPDF.
That speed gain is moderate in \CPP, more significant in Fortran and
Python.
As concerns Fortran, there are straightforward modifications to LHAPDF
that would improve its speed and these have been proposed to the
LHAPDF authors.

The PDF and $\alpha_s$ evaluation routines in \hoppet are
thread safe.
Other parts of the code, notably initialisation and evolution, are not.




\section{Conclusion}

Version 2 of \hoppet brings major additions to its functionality.
These include evolution up to N$^3$LO, massless structure function
evaluation, QED evolution, a Python interface, a modern build system,
functionality for writing LHAPDF grids and significant speed
improvements in the interpolation of its internal PDF tables.

Overall \hoppet remains highly competitive in terms of speed and
accuracy.
For example, repeated filling of a full PDF tabulation takes about a
millisecond per initial condition, with a relative accuracy of
$10^{-4}$ or better for $10^{-5} < x<0.9$.
It offers explicit handles to control the accuracy, allowing users to
verify the precision of their results and choose the optimal trade-off
between speed and precision.
Its modern Fortran interface also offers powerful and flexible access
to a range of common PDF manipulations such as convolutions with
arbitrary splitting and coefficient functions, features that are
useful in a variety of contexts.

We hope that this release of \hoppet can help provide solid
foundations for a range of groups to contribute to the ongoing
discussions~\cite{McGowan:2022nag,NNPDF:2024nan,Cooper-Sarkar:2024crx,Cridge:2024icl,Cooper-Sarkar:2025sqw}
in the field concerning the impact of \ntlo and QED effects in PDF
fits.
We also hope that the interfaces across computing languages will
facilitate the practical aspects of integration with a range of other
tools.

\section*{Acknowledgements}

We are grateful to Johannes Bl\"umlein for providing us with a
pre-release version of an exact Fortran code corresponding
Ref.~\cite{BlumleinCode} as well as a suitable license for its use,
and to Arnd Behring for assistance with \ttt{libome}.
We also gratefully acknowledge Luca Buonocore for his implementation
of the $P_{lq}$ splitting function in the QED code, Andrii Verbitskyi
for contributing the initial version of the CMake build system,
and Melissa van Beekveld for collaboration on initial options for speed
improvements in the evaluation of tabulated PDFs.
We thank Valerio Bertone for cross-checks of the  structure
functions and PDF evolution with \APFELPP.
We also wish to thank Juan Rojo for useful discussions. 
GPS acknowledges funding from a Royal Society Research
Professorship (grant RP$\backslash$R$\backslash$231001) and from the Science and
Technology Facilities Council (STFC) under grant ST/X000761/1.
PN thanks the Humboldt Foundation for support. 
%

\bibliographystyle{elsarticle-num}
\bibliography{hoppet.bib}

\end{document}